\documentclass[12pt,letterpaper]{article}
\usepackage{amssymb}
\usepackage{amsmath}
\usepackage{amsfonts}
\usepackage{epsfig}
\usepackage{feynmf}
\usepackage{color}
\usepackage{hyperref}

\usepackage{bm}
 \font\small=cmr10 scaled \magstep0
  
\outer\def\beginsection#1\par{\medbreak\bigskip
      \message{#1}\leftline{\bf#1}\nobreak\medskip
\vskip-\parskip
      \noindent}

\newcommand{\eq}{\begin{equation}}
\newcommand{\eqx}{\end{equation}}
\newcommand{\eqn}{\begin{eqnarray}}
\newcommand{\eqnx}{\end{eqnarray}}
\newcommand{\bi}{\begin{itemize}}
\newcommand{\ei}{\end{itemize}}

\def\Ord{{\cal O}}

\def\be{\begin{equation}}
\def\ee{\end{equation}}
\def\ba{\begin{eqnarray}}
\usepackage[titles]{tocloft}

\begin{document}
%%%%%%%%%%%%%%%%%%%%%%%
\begin{titlepage}
\hfill \hbox{NORDITA-2015-84}
\vskip 1.5cm
\vskip 1.0cm
\begin{center}
{\Large \bf  Double-soft behavior  for scalars and gluons  from   string  
theory }
 
\vskip 1.0cm {\large Paolo
Di Vecchia$^{a,b}$,
Raffaele Marotta$^{c}$, Matin Mojaza$^{b}$ } \\[0.7cm] 
{\it $^a$ The Niels Bohr Institute, University of Copenhagen, Blegdamsvej 17, \\
DK-2100 Copenhagen {\O}, Denmark}\\
{\it $^b$ Nordita, KTH Royal Institute of Technology and Stockholm University, \\Roslagstullsbacken 23, SE-10691 Stockholm, Sweden}\\
{\it $^c$  Instituto Nazionale di Fisica Nucleare, Sezione di Napoli, Complesso \\
 Universitario di Monte S. Angelo ed. 6, via Cintia, 80126, Napoli, Italy}
\end{center}
\begin{abstract}

We compute the leading double-soft behavior for gluons and for the scalars obtained
by dimensional reduction of a higher dimensional pure gauge theory, from the 
scattering amplitudes of gluons and scalars living in the world-volume of a D$p$-brane
of the bosonic string.  In the case of gluons, we  compute both the 
double-soft behavior when the two soft gluons are contiguous as well as when 
they are not contiguous.  From our results, that  are valid in string theory, one can easily
get the double-soft limit in gauge field theory by sending the string tension to infinity.

\end{abstract}
\end{titlepage}

%%%%%%%%%%%%%%%%%%%
\tableofcontents

\section{Introduction and results}
\label{intro}

Soft theorems were in the 1950s and 1960s studied for photons~\cite{LowFourPt, LowTheorem, Weinberg, 
OtherSoftPhotons} and gravitons~\cite{Weinberg,GrossJackiw,Jackiw},
when they were recognized to be 
important consequences of local gauge invariance~\footnote{For a discussion of 
low-energy theorem for photons see Chapter 3 of Ref.~\cite{DFFR}}. 
More recent discussions  of the generic subleading behavior of soft gluons and gravitons were given
in Refs.~\cite{WhiteYM,WhiteGrav}.
The very recent
interest in the soft behavior of gravitons and gluons  
has arisen after work by Strominger and
collaborators~\cite{Strominger}, showing that
the soft-graviton behavior follows from Ward identities of extended Bondi,
van der Burg, Metzner and Sachs (BMS) symmetry~\cite{BMS,ExtendedBMS},
and that tree-level graviton amplitudes in four spacetime dimensions have a universal behavior through
second subleading order in the soft-graviton momentum~\cite{CachazoStrominger}.
This has stimulated the study of the subleading soft 
 behavior in amplitudes with gluons and gravitons, which we will briefly summarize.
 
For gluons  in arbitrary number of  
dimensions soft theorems were obtained in various ways  at 
tree level~\cite{SoftGluonProof,Volovich,Conformal,IntegrandSoft,
TwistorSoft}, and discussed for quark-gluon amplitudes in QCD in Ref.~\cite{Luo:2014wea}.
Poincar{\'{e}} and gauge invariance as well as a condition arising from the 
distributional nature of scattering amplitudes have been used  in 
Ref.~\cite{NewPaper} to  strongly constrain the soft behavior for gluons 
and gravitons, while in Ref.~\cite{BDDN} gauge invariance is shown 
to completely fix the first two   
leading terms  (up to terms $\Ord (q^0)$) in the case of a gluon, and
 the first three leading terms (up to terms $\Ord (q^1)$) in the case 
of a graviton, for any number 
 of space-time dimensions ($q$ being the soft momentum).

Further study   of the  subleading soft-graviton
 theorems in  arbitrary number of  
dimensions were performed  in 
Refs.~\cite{Royas,Zlotnikov,LNS,CHW,SabioVera,
White:2014qia,Du:2014eca}. Soft theorems were also studied 
in the ambitwistor string~\cite{Lipstein:2015rxa,Casali:2015vta},
and their connection with effective field theories was discussed in 
Ref.~\cite{Cheung:2014dqa}.
As discussed in Refs.~\cite{LoopCorrections,HeHuang,BDDN,DDLPR},
one should note that soft gluon and graviton behaviors are in general 
modified by loop corrections.

Finally, soft gluon and graviton behavior have also been studied 
in the framework of string theory in 
Refs.~\cite{IntegrandSoft,StringSoft,Schwab,DiVecchia:2015oba}.  In particular, soft
theorems for the 
dilaton were discussed in Refs.~\cite{Ademollo:1975pf,Shapiro:1975cz,Yoneya:1987gc,Hata:1992it,
DiVecchia:2015oba} 
and for the anti-symmetric tensor in Ref.~\cite{DiVecchia:2015oba}, while soft behavior in amplitudes
with massive states was discussed in Ref.~\cite{1505.05854}.

Double-soft theorems are now receiving increasing interest, and have been studied
for scalars and gluons. The interest in the double-soft behavior arises originally 
from the analysis  of the spontaneously symmetry breaking of a group $G$ 
keeping  a subgroup $H$ unbroken. The  presence of  Nambu-Goldstone 
bosons,  living in  the coset space $G/H$, implies  the
vanishing of the scattering amplitude with a single soft  Goldstone 
boson, giving rise to the famous Adler's zeros, while leading to 
 a peculiar nonzero universal behaviour of the amplitude with
two soft Goldstone  particles\cite{ArkaniHamed:2008gz, 1412.2145}.
The double-soft theorems for different kinds of scalars were
computed in Ref.~\cite{Cachazo:2015ksa} using the CHY representation 
of  the scattering  amplitude~\cite{Cachazo:2013hca}. 
The double-soft theorems for gluons were computed in 
Refs.~\cite{Klose:2015xoa,Volovich:2015yoa} in four dimensions 
using the spinor  helicity formalism and  the BCFW recursion relations, and
this result has since been confirmed in Ref.~\cite{Volovich:2015yoa}  by a calculation 
in an arbitrary  number of spacetime dimensions,
using the CHY representation of the scattering amplitude. 
Ref.~\cite{Volovich:2015yoa} also uses superstring theory obtaining agreement
with field theory results without extra string corrections. 
Finally, in Ref.~\cite{Georgiou:2015jfa} 
the so-called CSW method is used to compute the double-soft behavior in four 
dimensions.

In this work we compute the leading double-soft limit for gluons and scalars 
directly on  the scattering amplitudes for gluons 
and scalars in  the  bosonic string. The corresponding result in the limiting field theory
is then obtained by performing the field theory limit ($\alpha' \rightarrow 0$).
 Our results are valid for any number 
of space-time dimensions   and without 
fixing a particular gauge.
In particular, we consider  scattering amplitudes of  massless  open  
strings on a D$p$-brane  in the bosonic string 
 and use them  to derive  double-soft theorems for
both gauge fields  and scalars. In fact, in the presence of a  D$p$-brane, the original 
Lorentz symmetry $SO(1, d-1)$ is broken into $SO(1,p) \otimes SO(d-1-p)$
and the original $d$-dimensional gauge field  gives rise to a 
$(p+1)$-dimensional gauge field that we in the following, for the sake of simplicity,
  call gluon, 
and $d-p-1$ massless scalars~\footnote{Here and in the following we keep $d$ arbitrary
although for the bosonic string we need to take $d=26$.}. They all live 
on the $(p+1)$-dimensional world-volume of the D$p$-brane.
 
We start by computing the color-ordered amplitude with $(n+2)$ gluons. 
We keep 
only the leading  term   when  two contiguous gluons with momenta $q_1$ and $q_2$
 become simultaneously soft,  while the momenta $k_1 \dots k_n$   of the other gluons 
 are  kept finite.   In this case, where the gluons are ordered as 
$k_1 \dots  k_{n-1}\,  q_1\,  q_2 \, k_n$, we find that the leading 
term  in  
 the double-soft limit  behaves as  $\frac{1}{q_{1,2}^2}$  and is given by:
\begin{eqnarray}
M_{2g;ng}   = 
\frac{g_{p+1}^{2} }{ q_1 q_2}  
 &&\left\{ 
- \frac{1 }{2}
 (\epsilon_{q_1} \epsilon_{q_2} )   \left[ \frac{k_n (q_2 - q_1) + 
q_1 q_2 }{k_n (q_1 + q_2) + q_1 q_2  }  + \frac{k_{n-1} (q_1 - q_2) + 
q_1 q_2 }{k_{n-1} (q_1 + q_2) + q_1 q_2  }   \right]  \right. \nonumber \\
&& + 
\frac{ (\epsilon_{q_1} q_2 ) (\epsilon_{q_2} k_n) -  (\epsilon_{q_2} q_1 ) 
(\epsilon_{q_1} k_n)}{  ( k_n (q_1 + q_2) + q_1 q_2 ) } 
-  \frac{ (\epsilon_{q_1} q_2 ) (\epsilon_{q_2} k_{n-1}) -  (\epsilon_{q_2} q_1 ) (\epsilon_{q_1} k_{n-1}
)}{  ( k_{n-1} (q_1 + q_2) + q_1 q_2 ) }  \nonumber \\
&&   + \frac{(\epsilon_{q_1} k_n) (\epsilon_{q_2} k_n) (q_1 q_2)}{(k_n q_2) ( k_n (q_1 + q_2) +
 q_1 q_2 )} +  \frac{(\epsilon_{q_1} k_{n-1}) (\epsilon_{q_2} k_{n-1}) (q_1 q_2)}{(k_{n-1} q_1)
 ( k_{n-1} (q_1 + q_2) + q_1 q_2 )} 
\nonumber \\
&& \left. - \frac{(\epsilon_{q_1} k_{n-1}) ( \epsilon_{q_2} k_n) (q_1 q_2) }{(k_{n-1} q_1 ) (k_n q_2)}
 \right\}  M_{ng} \ , \nonumber \\
\label{nowfff}
\end{eqnarray}
where $g_{p+1}$ is the gauge coupling constant of the gauge theory 
living on the D$p$-brane and   
\begin{eqnarray}
&& M_{ng}  =  \frac{(2\alpha')^{\frac{n-2}{2}} }{\alpha'} g_{p+1}^{n-2}  \int
 \frac{ \prod_{i=1}^{n} dz_i }{dV_{abc}} 
\prod_{i=1}^{n} d \theta_i \langle 0 | 
 \prod_{i=1}^{n} {\rm e}^{i \left( \theta_i 
\epsilon_i \partial_{z_i} +
\sqrt{2\alpha'} k_{i}  \right) X (z_i) } |0 \rangle  
 \label{Annnnnintro}
\end{eqnarray}
is the scattering amplitude of $n$ gluons in the bosonic string.
As we explicitly show in the next sections the
dependence on $\alpha'$ in the soft factor drops out, while it is still present 
in the $n$-gluon amplitude. The field-theory limit, corresponding to  $\alpha' \rightarrow 0$, 
leaves the soft factor unchanged and acts only on $M_{ng}$ giving
 then just the $n$-gluon amplitude of Yang-Mills theory. Notice also 
that, unlike the case of a single
soft behavior, the double-soft behavior cannot be written as the difference of two terms, one
depending on $k_{n-1}$ and the other on $k_n$, because of the presence of the last term
in Eq. (\ref{nowfff}) that contains both momenta.

We then consider the double-soft limit for two identical scalars in an amplitude with $n$ gluons
and we get the following double-soft behavior:
\begin{eqnarray}
M_{2s;ng} = -\frac{g_{p+1}^{2}}{2q_1 q_2} \left[\frac{k_n (q_2 - q_1) + q_1 q_2}{k_{n}
 (q_1 + q_2) + q_1  q_2} + \frac{ k_{n-1} (q_1 - q_2) + q_1 q_2}{k_{n-1}
 (q_1 + q_2) + q_1  q_2}   \right] M_{ng}
\label{Mng2sofinalnn}
\end{eqnarray}
where $M_{ng}$ is again given in Eq. (\ref{Annnnnintro}).
Finally, we get the same double-soft behavior for a scattering amplitude with $n+2$ identical 
scalars when two contiguous scalars  become soft:
\begin{eqnarray}
M_{2s;ns} =  -\frac{g_{p+1}^{2}}{2q_1 q_2} \left[ \frac{k_n (q_2 - q_1) + q_1 q_2}{k_{n}
 (q_1 + q_2) + q_1  q_2} + \frac{ k_{n-1} (q_1 - q_2) + q_1 q_2}{k_{n-1}
 (q_1 + q_2) + q_1  q_2}   \right] M_{ns}
\label{M2nsnsxxx}
\end{eqnarray}
where $M_{ns}$ is the $n$-scalar amplitude in the bosonic string.

The three previous formulas are valid in an arbitrary $(p+1)$-dimensional space-time and all
follow from a $d$-dimensional gauge theory that, because of the presence of a D$p$-brane,
gives rise to a gauge theory in a $(p+1)$-dimensional space-time coupled to $(d-p-1)$ scalar 
fields, which correspond to the components of the original $d$-dimensional gauge field along
the directions outside the world-volume of the D$p$-brane. Because of this, 
Eqs. (\ref{Mng2sofinalnn}) and (\ref{M2nsnsxxx}) follow from Eq. (\ref{nowfff}) by neglecting 
the terms where the polarizations of the soft gluons are contracted with the momenta
of all particles. This is a consequence of  the fact that the scalars correspond to the components
of the gluons in the directions orthogonal to the D$p$-brane.

It is also interesting to note that the previous  three soft amplitudes all involve the factor  in 
the squared bracket of Eqs. 
(\ref{Mng2sofinalnn}) and (\ref{M2nsnsxxx}) that is typical of the double-soft limit of 
Goldstone bosons~\cite{ArkaniHamed:2008gz}.  Unlike the case of 
 Goldstone bosons, however, in this case  
we get  an additional singular factor  $\frac{1}{q_1\cdot q_2}$. 

We have also computed the leading double-soft behavior of a color-ordered 
 amplitude with $n+2$ gluons
in the case where the two soft gluons are not next to each other as before, but 
have a hard gluon between them, which we take to be the one with momentum $k_{n-1}$. 
When the gluons are ordered as $k_1 , k_2 \dots q_1 \,k_{n-1} \,q_2 \,k_n$,  the leading 
double-soft behavior is given by
\begin{eqnarray}
M_{2g;ng} =   g_{p+1}^2 && \left[ \frac{ k_{n-2} \epsilon_{q_1}}{k_{n-2} q_1 } 
\left(  \frac{k_{n-1} \epsilon_{q_2}}{k_{n-1} q_2} -  \frac{k_{n} 
\epsilon_{q_2}}{k_{n} q_2} \right) +  \frac{k_{n-1} \epsilon_{q_1}}{k_{n-1} q_1}
 \frac{k_{n} \epsilon_{q_2}}{k_{n} q_2}  \right. \nonumber \\
 && \left.  -  \frac{ (\epsilon_{q_1} k_{n-1})  (\epsilon_{q_2} k_{n-1}) }{ k_{n-1}(q_1 + q_2) + 
q_1 q_2   } \left( \frac{1}{ k_{n-1} q_1} +
 \frac{1}{ k_{n-1} q_2}   \right)  \right]M_{ng}
\label{sidste} 
\end{eqnarray}  
Finally, for the color-ordered amplitude where the  two soft gluons have more than one 
hard gluon between them, the double-soft behavior is just given by the product of two 
single-soft behaviors.

The paper is organized as follows. In Sect. \ref{double-soft-with-gluons} we discuss the 
double-soft limit for gluons in an amplitude with $n$ gluons. Sect. 
\ref{double-soft-with-scalars-gluon} is devoted to the double-soft limit for two scalars in 
an amplitude
with $n$ gluons, while in Sect. \ref{double-soft-with-scalars} we discuss the leading
double-soft limit for two scalars in an amplitude with $n$ scalars.  The last section is left
for conclusions. In the Appendix we compute the double-soft limit of some double integrals
used for computing the double-soft limits.

\section{Double-soft behavior with $n+2$ gluons}
\setcounter{equation}{0}
\label{double-soft-with-gluons}

In this section we consider the color-ordered scattering amplitude involving 
$(n+2)$ gauge fields living on the world-volume of a D$p$-brane 
of the bosonic string and we compute the leading double-soft behavior
when two contiguous  gluons become simultaneously soft. 

We denote with $(\epsilon_{q_1} , q_1)$ and  $(\epsilon_{q_2} , q_2)$  
the polarizations and momenta of the gluons that eventually will become
soft and with $(\epsilon_i , k_i)$ the polarizations and 
momenta of the remaining gluons. We consider the color-ordered 
amplitude corresponding to the following permutation: $ k_1 , k_2, \dots
k_{n-1}, q_1 , q_2 , k_n$ for which the Koba-Nielsen variables of the various gluons are 
ordered as $ z_1 \geq z_2 \geq z_3 \dots \geq z_{n-1} \geq w_1 \geq w_2
\geq z_n$. 

It is convenient to write the amplitude with $(n+2)$ gluons by exponentiating
the derivative part of the  vertex operators by introducing for each 
vertex operator a Grassmann variable, called $\theta_i , i=1 \dots n$ 
and $\phi_a , a=1,2$,
and by writing the scattering amplitude as follows~\footnote{
Here and in the following we  assume that $z_n =0$.}:
\begin{eqnarray}
M_{2g;ng} = && \frac{(\sqrt{2 \alpha'})^{n}}{\alpha'  } g_{p+1}^{n}
\int \frac{ \prod_{i=1}^{n} dz_i }{dV_{abc}} \prod_{i=1}^{n} d \theta_i 
  \int_{0}^{z_{n-1}}  dw_1 \int_{0}^{w_1}
dw_2 \int d\phi_1 d \phi_2 \nonumber \\
&& \times 
\langle 0 |  \prod_{i=1}^{n-1} {\rm e}^{i \left( \theta_i 
\epsilon_i  \partial_{z_i} +
\sqrt{2\alpha'} k_{i}   \right) X  (z_i) }     \label{original} \\
&&  \times {\rm e}^{i \left( \phi_1 \epsilon_{q_1}   \partial_{w_1} +
\sqrt{2\alpha'} q_{1}   \right) X  (w_1) }   {\rm e}^{i \left( \phi_2
 \epsilon_{q_2}  \partial_{w_2} +\sqrt{2\alpha'} q_{2}  \right)
 X  (w_2) }  {\rm e}^{i \left( \theta_n \epsilon_n \partial_{z_n} +
\sqrt{2\alpha'} k_{n}  \right) X  (z_n) } \rangle \ , \nonumber
\end{eqnarray}
where $g_{p+1}$ is the $(p+1)$-dimensional gauge coupling constant~\footnote{In order to 
have
a bosonic quantity in the exponents, we assume that also the polarization 
vectors are Grassmann
variables.}.
Using the contraction
\begin{eqnarray}
X_{\mu} (z) X_{\nu} (w) \sim - \eta_{\mu \nu} \log (z-w)~~\text{ with }~~\eta_{\mu \nu} =
(- + + \dots +)
\label{contra}
\end{eqnarray}
we can contract the vertex operators  of the 
states with momenta $q_1$ and $q_2$, with those of the other states getting
\begin{eqnarray}
 M_{2g;ng} = && \frac{(\sqrt{2 \alpha'})^{n}}{\alpha'  } g_{p+1}^{n} 
\int \frac{ \prod_{i=1}^{n} dz_i }{dV_{abc}} \prod_{i=1}^{n} d \theta_i 
\langle 0 |  \prod_{i=1}^{n} {\rm e}^{i \left( \theta_i \epsilon_i
\partial_{z_i} +\sqrt{2\alpha'} k_{i}  \right) X (z_i) } |0 \rangle
\nonumber \\
&& \times 
  \int_{0}^{z_{n-1}}  dw_1 \int_{0}^{w_1}
dw_2 \int d\phi_1 d \phi_2 \nonumber \\
&& \times \prod_{i=1}^{n-1} \left( (z_i - w_1)^{2\alpha' k_i q_1} (z_i -
 w_2)^{2\alpha' k_i q_2} \right)
(w_1 - z_n )^{2\alpha' q_1 k_n} (w_2 - z_n )^{2\alpha' q_2 k_n} \nonumber \\
&& \times (w_1- w_2)^{2\alpha' q_1 q_2} \prod_{i=1}^{n} \left(
 {\rm e}^{\sqrt{2\alpha'}  
\frac{\theta_i \epsilon_i q_1}{z_i - w_1}} {\rm e}^{\sqrt{2\alpha'}  \frac{\theta_i 
\epsilon_i q_2}{z_i - w_2}} \right)
 {\rm e}^{- \phi_1 \phi_2 \frac{ (\epsilon_{q_1} 
\epsilon_{q_2})}{(w_1 - w_2)^2}}\nonumber \\
&& \times {\rm e}^{ \phi_1 \left[  \sum_{i=1}^{n} \frac{\theta_i 
( \epsilon_{q_1} \epsilon_i )}{(z_i - w_1)^2}  - \sum_{i=1}^{n} 
 \frac{\sqrt{2\alpha'} (k_i 
\epsilon_{q_1} )}{z_i - w_1}      + \frac{\sqrt{2\alpha'} (
\epsilon_{q_1} q_2 )}{w_1 - w_2}    \right] } \nonumber \\
&& \times {\rm e}^{ \phi_2 \left[  \sum_{i=1}^{n}  
\frac{\theta_i ( \epsilon_{q_2} \epsilon_i )}{(z_i - w_2)^2}  - \sum_{i=1}^{n} 
\frac{\sqrt{2\alpha'} (k_i \epsilon_{q_2} )}{z_i - w_2}  -
 \frac{\sqrt{2\alpha'} (\epsilon_{q_2} q_1 )}{w_1 - w_2}    \right] } \ .
\label{n+2gluon}
\end{eqnarray}
We can now perform the  Grassmann integrals over $\phi_1$ and $\phi_2$ arriving
at
\begin{eqnarray}
M_{2g;ng}= && 
\frac{(2\alpha')^{\frac{n-2}{2}} }{\alpha'} g_{p+1}^{n-2}  \int
 \frac{ \prod_{i=1}^{n} dz_i }{dV_{abc}} 
\prod_{i=1}^{n} d \theta_i \langle 0 | 
 \prod_{i=1}^{n} {\rm e}^{i \left( \theta_i 
\epsilon_i \partial_{z_i} +
\sqrt{2\alpha'} k_{i} \right) X (z_i) } |0 \rangle  
 \nonumber \\
&& \times   (2 \alpha' g_{p+1}^2)  \int_{0}^{z_{n-1}} dw_1 
 \int_{0}^{w_1} dw_2  
\prod_{i=1}^{n-1} \left( (z_i - w_1)^{2\alpha' k_i q_1} (z_i - w_2)^{2\alpha'
 k_i q_2} \right) \nonumber \\
&& \times (w_1 - z_n )^{2\alpha' q_1 k_n} (w_2 - z_n )^{2\alpha' q_2 k_n} 
 (w_1- w_2)^{2\alpha' q_1 q_2} \prod_{i=1}^{n} \left( {\rm e}^{\sqrt{2\alpha'} 
 \frac{\theta_i \epsilon_i q_1}{z_i - w_1}} {\rm e}^{\sqrt{2\alpha'}  \frac{\theta_i
 \epsilon_i q_2}{z_i - w_2}} \right)
  \nonumber \\
 &&\times  \left\{   \frac{ (\epsilon_{q_1} \epsilon_{q_2})}{(w_1 - w_2)^2}   
+ 
 \left[  \sum_{i=1}^{n}  \frac{\theta_i ( \epsilon_i  \epsilon_{q_1} 
)}{(z_i - w_1)^2}  - \sum_{i=1}^{n} \frac{\sqrt{2\alpha'} (k_i 
\epsilon_{q_1} )}{z_i - w_1}  
   + \frac{\sqrt{2\alpha'} (\epsilon_{q_1} q_2 )}{w_1 - w_2}    \right]  \right.
\nonumber \\
&& \left.  \times  \left[  \sum_{j=1}^{n} \frac{\theta_j ( 
\epsilon_j \epsilon_{q_2}  )}{(z_j - w_2)^2}  - \sum_{j=1}^{n} 
\frac{\sqrt{2\alpha'} (k_j \epsilon_{q_2} )}{z_j - w_2}  
  - \frac{\sqrt{2\alpha'} (\epsilon_{q_2} q_1 )}{w_1 - w_2}    \right]  \right\}
\equiv  M_{ng} * G_n \ ,
\label{soft2gluons}
\end{eqnarray}
where  by $*$ a convolution of the integrals is understood,
\begin{eqnarray}
&& M_{ng}  =  \frac{(2\alpha')^{\frac{n-2}{2}} }{\alpha'} g_{p+1}^{n-2}  \int
 \frac{ \prod_{i=1}^{n} dz_i }{dV_{abc}} 
\prod_{i=1}^{n} d \theta_i \langle 0 | 
 \prod_{i=1}^{n} {\rm e}^{i \left( \theta_i 
\epsilon_i \partial_{z_i} +
\sqrt{2\alpha'} k_{i}  \right) X (z_i) } |0 \rangle  
 \label{Annnnn}
\end{eqnarray}
is the scattering amplitude of $n$ gluons,
and
\begin{eqnarray}
G_n = &&   (2 \alpha' g_{p+1}^2)  \int_{0}^{z_{n-1}} dw_1  \int_{0}^{w_1} dw_2  
\prod_{i=1}^{n} \left( (z_i - w_1)^{2\alpha' k_i q_1} (z_i - w_2)^{2\alpha' k_i q_2} 
\right) \nonumber \\
&& \times
 (w_1- w_2)^{2\alpha' q_1 q_2}  \prod_{i=1}^{n} 
\left( {\rm e}^{\sqrt{2\alpha'}  \frac{\theta_i \epsilon_i q_1}{z_i - w_1}}
 {\rm e}^{\sqrt{2\alpha'}  \frac{\theta_i \epsilon_i q_2}{z_i - w_2}} \right)
 \nonumber \\
  &&\times  \left\{   \frac{ (\epsilon_{q_1} \epsilon_{q_2})}{(w_1 - w_2)^2}   
+ 
 \left[ \sum_{i=1}^{n} \frac{\theta_i ( \epsilon_i  \epsilon_{q_1} )}{(z_i - w_1)^2} 
 - \sum_{i=1}^{n} \frac{\sqrt{2\alpha'} (k_i \epsilon_{q_1} )}{z_i - w_1}  
  + 
\frac{\sqrt{2\alpha'} (\epsilon_{q_1} q_2 )}{w_1 - w_2}    \right] \right.
 \nonumber \\
&& \left.  \times  \left[ \sum_{j=1}^{n} \frac{\theta_j 
( \epsilon_j  \epsilon_{q_2} )}{(z_j - w_2)^2}  - \sum_{j=1}^{n}
 \frac{\sqrt{2\alpha'} (k_j \epsilon_{q_2} )}{z_j - w_2}       -
 \frac{\sqrt{2\alpha'} (\epsilon_{q_2} q_1 )}{w_1 - w_2}    \right]  \right\} \ .
\label{softfactorSbis}
\end{eqnarray}
This expression contains two kinds of terms. The first one is without any dependence on the
variables $\theta_i$  and  acts in the convolution at the end  of  Eq.~(\ref{soft2gluons}) as 
a factor that multiplies the $n$-gluon 
amplitude $M_{ng}$, while the second one contains the terms with $\theta_i$ that, when acting
in  the convolution in Eq. (\ref{soft2gluons}), modify  the structure of the $n$-gluon 
amplitude $M_{ng}$. It can be shown that the first kind of terms is the leading 
one in the double-soft limit,
while the second one is subleading. In this paper we compute only the leading 
one that is equal to
\begin{eqnarray}
 G_n^{(1)}=  &&
g_{p+1}^2(2\alpha')~ \int_0^{z_{n-1}}dw_1\int_0^{w_1}dw_2~(w_1-w_2)^{2\alpha' q_1
\cdot q_2}~~w_1^{2\alpha' q_1\cdot k_n}~w_2^{2\alpha' q_2\cdot k_n}\nonumber\\
 &&\times~ (z_{n-1}-w_1)^{2\alpha' 
q_1\cdot k_{n-1}} (z_{n-1}-w_2)^{2\alpha' q_2\cdot k_{n-1}} 
\nonumber \\
&& \times \biggl\{ \frac{\epsilon_{q_1}\cdot 
\epsilon_{q_2}}{(w_1-w_2)^2}+ \biggl[  
\frac{\sqrt{2\alpha'} (k_{n-1} \epsilon_{q_1})}{w_1-z_{n-1}} 
+ \frac{\sqrt{2\alpha'} (k_n \epsilon_{q_1})
}{w_1}
+\frac{\sqrt{2\alpha'} (q_2 \epsilon_{q_1})}{{w}_1-{w}_2}  \biggr] 
\nonumber \\
&& \times
\biggl[  
\frac{\sqrt{2\alpha'} (k_{n-1} \epsilon_{q_2}) }{w_2-z_{n-1}}+ \frac{\sqrt{2\alpha'} (k_n \epsilon_{q_2})
}{w_2}+
\frac{\sqrt{2\alpha'} (q_1 \epsilon_{q_2})}{w_2-w_1} \biggr]\biggr\}
\label{Snalmostcomple}
\end{eqnarray} 
It is convenient to introduce two new variables $z_{n-1} {\hat{w}}_{a} = w_{a}$ 
for $a=1,2$.
We get
\begin{eqnarray}
 G_n^{(1)}=  &&
g_{p+1}^2(2\alpha')~ z_{n-1}^{ 2\alpha' [q_1 q_2 +(k_{n-1} + k_n) (q_1 + q_2)
  ]}  
\int_0^{1}}d{{\hat{w}}_1\int_0^{ {\hat{w}}_1}
d {\hat{w}}_2~({\hat{w}}_1-{\hat{w}}_2)^{2\alpha' q_1
\cdot q_2}  \nonumber \\
&& \times {\hat{w}}_1^{2\alpha' q_1\cdot k_n}~{\hat{w}}_2^{2\alpha' 
q_2\cdot k_n}
~ (1-{\hat{w}}_1)^{2\alpha' 
q_1\cdot k_{n-1}} (1-{\hat{w}}_2)^{2\alpha' q_2\cdot k_{n-1}} 
\nonumber \\
&& \times \biggl\{ \frac{\epsilon_{q_1}\cdot 
\epsilon_{q_2}}{({\hat{w}}_1-{\hat{w}}_2)^2}+ \biggl[  
- \frac{\sqrt{2\alpha'} (k_{n-1} \epsilon_{q_1})}{ 1 - {\hat{w}}_1} 
+ \frac{\sqrt{2\alpha'} (k_n \epsilon_{q_1})
}{{\hat{w}}_1}
+\frac{\sqrt{2\alpha'} (q_2 \epsilon_{q_1})}{{{\hat{w}}}_1-{\hat{w}}_2}  \biggr] 
\nonumber \\
&& \times
\biggl[  -
\frac{\sqrt{2\alpha'} (k_{n-1} \epsilon_{q_2}) }{1- {\hat{w}}_2}+ \frac{\sqrt{2\alpha'} (k_n 
\epsilon_{q_2})
}{{\hat{w}}_2}-
\frac{\sqrt{2\alpha'} (q_1 \epsilon_{q_2})}{{\hat{w}}_1-{\hat{w}}_2} \biggr]\biggr\}
\label{Snalmostcomple1}
\end{eqnarray} 
We want to extract from the previous sum of integrals 
the most singular term in the double-soft limit.
In this limit, we can approximate 
$z_{n-1}^{ 2\alpha' [q_1 q_2 +(k_{n-1} + k_n) (q_1 + q_2)  ]}$ with $1$.  
The double-soft behavior of the various integrals is computed in 
Appendix \ref{computation}.  Here we give only 
the final result:
\begin{eqnarray}
 G_{n}^{(1)}= && g_{p+1}^{2} \bigg\{-\frac{(\epsilon_{q_1} \epsilon_{q_2})}{2q_1
 q_2}\biggl[ 
 \frac{k_n  (q_2-q_1)+q_1  q_2}{k_n (q_1+q_2)+q_1 q_2} +
 \frac{k_{n-1} (q_1-q_2)+q_1  q_2}{k_{n-1} (q_1+q_2)+q_1
 q_2}\biggr] +
\nonumber\\
 && \times \frac{1}{q_1  q_2}\biggl[\frac{(\epsilon_{q_1}  
q_2) ~ (\epsilon_{q_2}  k_n)-\
(\epsilon_{q_2}  q_1)~(\epsilon_{q_1}  k_n)}{k_n 
(q_1+q_2)+q_1 q_2} - 
\frac{ (\epsilon_{q_1}  q_2)~ (\epsilon_{q_2}  k_{n-1}) - 
(\epsilon_{q_2}  q_1)~ 
(\epsilon_{q_1}  k_{n-1})}{k_{n-1}  (q_1+q_2)+q_1
 q_2}\biggr]\nonumber\\
 &&+\frac{(\epsilon_{q_1}  k_n)~(\epsilon_{q_2}  k_n)}{q_2
 k_n~[k_n (q_1+q_2)
+q_1 q_2]}+\frac{( \epsilon_{q_1}  k_{n-1})~(\epsilon_{q_2}
 k_{n-1})}{q_1
 k_{n-1}~[k_{n-1} (q_1+q_2)+q_1\cdot q_2]} \nonumber \\
&& -\frac{(\epsilon_{q_1}  k_{n-1})~(\epsilon_{q_2}  k_n)}{(q_1
 k_{n-1})~(q_2 
k_n )} \bigg\} \ .
\label{Gn(1)}
\end{eqnarray} 
In conclusion,  the  leading double-soft behavior  is given by:
\begin{eqnarray}
M_{2g;ng}   =  
\frac{g_{p+1}^{2} }{ q_1 q_2}  
&& \left\{
- \frac{1 }{2}
 (\epsilon_{q_1} \epsilon_{q_2} )   \left[ \frac{k_n (q_2 - q_1) + 
q_1 q_2 }{k_n (q_1 + q_2) + q_1 q_2  }  + \frac{k_{n-1} (q_1 - q_2) + 
q_1 q_2 }{k_{n-1} (q_1 + q_2) + q_1 q_2  }   \right]  \right. \nonumber \\
&& + 
\frac{ (\epsilon_{q_1} q_2 ) (\epsilon_{q_2} k_n) -  (\epsilon_{q_2} q_1 ) 
(\epsilon_{q_1} k_n)}{  ( k_n (q_1 + q_2) + q_1 q_2 ) } 
-  \frac{ (\epsilon_{q_1} q_2 ) (\epsilon_{q_2} k_{n-1}) -  (\epsilon_{q_2} q_1 )
 (\epsilon_{q_1} k_{n-1}
)}{  ( k_{n-1} (q_1 + q_2) + q_1 q_2 ) }   \nonumber \\
&&   + \frac{(\epsilon_{q_1} k_n) (\epsilon_{q_2} k_n) (q_1 q_2)}{(k_n q_2) 
( k_n (q_1 + q_2) +
 q_1 q_2 )} +  \frac{(\epsilon_{q_1} k_{n-1}) (\epsilon_{q_2} k_{n-1}) 
(q_1 q_2)}{(k_{n-1} q_1)
 ( k_{n-1} (q_1 + q_2) + q_1 q_2 )} 
\nonumber \\
&& \left. - \frac{(\epsilon_{q_1} k_{n-1}) ( \epsilon_{q_2} k_n) 
(q_1 q_2) }{(k_{n-1} q_1 ) (k_n q_2)}
 \right\}  M_{ng} \ , \nonumber \\
\label{now}
\end{eqnarray}
which behaves as $\frac{1}{q_{1} q_2}$ in the double-soft limit,  i.e. when both $q_1$ and 
$q_2$ simultaneously go to zero. $M_{ng}$ is the $n$-gluon amplitude in the bosonic string.
We see that, in the double-soft limit, the dependence on $\alpha'$ of  the soft factor in Eq.
(\ref{Gn(1)} )  cancels, while it is still kept in the $n$-gluon amplitude in Eq. (\ref{Annnnn}).

One can check gauge invariance of the  soft factor  in Eq.
(\ref{Gn(1)} )  by substituting 
$\epsilon_{q_1}$ with $q_1$. One gets:
\begin{eqnarray}
- g_{p+1}^2 \left(   \frac{ (q_1 \epsilon_{q_2})}{k_{n-1} (q_1 + q_2) + q_1 q_2}  + 
\frac{(q_1 q_2) (\epsilon_{q_2} k_n)}{(k_n q_2)    
( k_n (q_1 + q_2) + q_1 q_2)}\right) \sim q_{1,2}^{0} \ .
\label{gauinber}
\end{eqnarray}
The original amplitude behaves as $\frac{1}{q_{1,2}^2}$. Saturating with $q_1$, we find that 
the terms  of order $\frac{1}{q_{1,2}}$ cancel and  we are left with a term of 
order $q_{1,2}^0$ that should be cancelled by the next to the leading term, which we have not yet computed.

One can also check gauge invariance with respect to the other soft particle by substituting
$\epsilon_{q_2}$ with $q_2$. One gets:
\begin{eqnarray}
- g_{p+1}^{2} \left( \frac{ (\epsilon_{q_1} q_2)  }{k_n (q_1 +q_2) + q_1 q_2 } +
  \frac{(\epsilon_{q_1} k_{n-1}) (q_1 q_2)}{( k_{n-1} (q_1 + q_2) + q_1 q_2 ) (k_{n-1} q_1)} \right) \ ,
\label{gsir}
\end{eqnarray}
which is equal to the expression in Eq. (\ref{gauinber}) by the substitutions 
$k_{n-1} \leftrightarrow k_n$ and $q_1 \leftrightarrow q_2$.

\begin{figure}[b]
\begin{center}
\includegraphics[width=.7\textwidth]{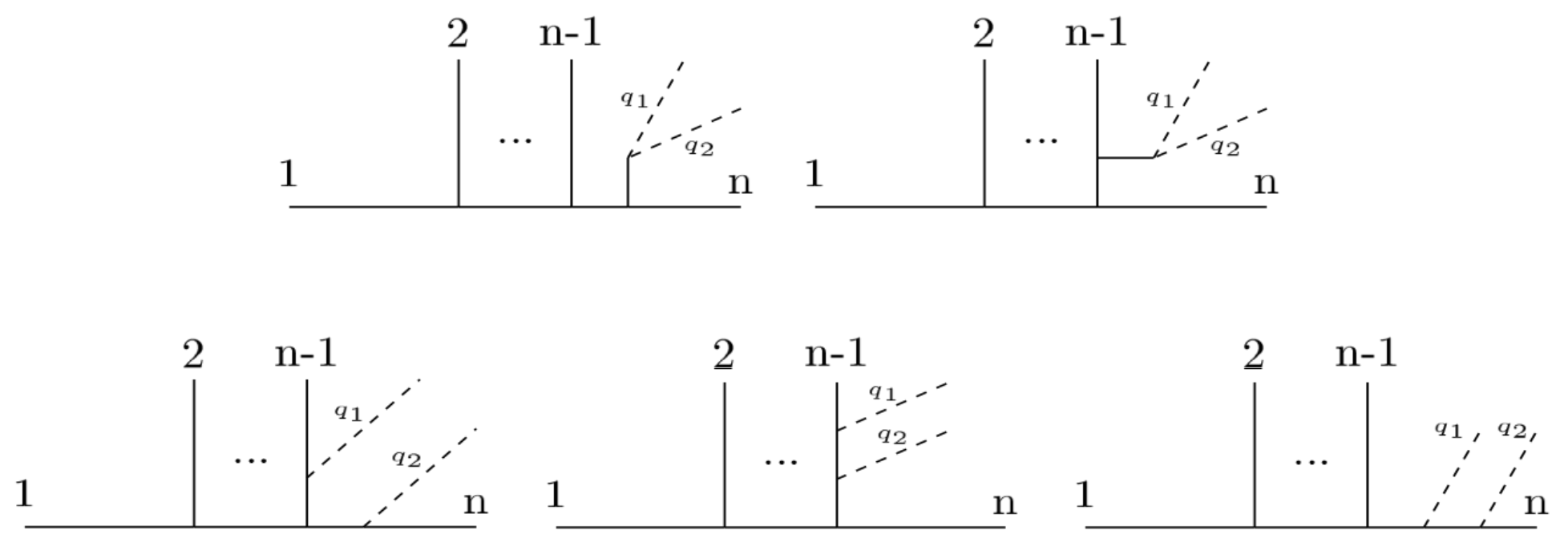}
\caption{ \small Sketchy diagrams of the color-ordered n+2 gluon amplitude,
where solid lines are the hard gluons and dashed lines are the soft gluons.
Each term in Eq.~(\ref{now12}) corresponds to one of the diagrams, and shows
that all poles allowed in the color-ordered amplitude contribute to the
double-soft limit.}
\label{Diagrams}
\end{center}
\end{figure}

In order to see the pole structure of the amplitude, it is convenient to 
write Eq. (\ref{now}) as follows:
\begin{eqnarray}
M_{2g;ng}   =  
&&\frac{2g_{p+1}^{2} }{ (q_1 + q_2)^2} \nonumber \\
&&\left\{  \left[ -
 (\epsilon_{q_1} \epsilon_{q_2} )    \frac{ k_n (q_2 - q_1) +q_1 q_2}{ 
  (k_n + q_1 + q_2)^2   } +   2
 \frac{ (\epsilon_{q_1} q_2 ) (\epsilon_{q_2} k_n) -  
(\epsilon_{q_2} q_1 ) (\epsilon_{q_1} k_n)}{ ( k_n +q_1 + q_2)^2  } \right. \right. 
 \nonumber \\ 
&&  \left.  + 2 \frac{ (\epsilon_{q_1} k_n) (\epsilon_{q_2} k_n) (q_1 + 
q_2)^2}{(k_n + q_2)^2 ( k_n 
+q_1 + q_2)^2 } \right] \nonumber \\ 
&&  -  \left[ - (\epsilon_{q_2} \epsilon_{q_1} )   \frac{k_{n-1} (q_1 - q_2) + 
q_1 q_2 }{ (k_{n-1} +q_1 + q_2)^2   }  +  2\frac{ (\epsilon_{q_1} q_2 )
 (\epsilon_{q_2} k_{n-1}) - 
 (\epsilon_{q_2} q_1 ) (\epsilon_{q_1} k_{n-1})}{  ( k_{n-1} +q_1 + q_2)^2  } \right.
  \nonumber \\
&& \left. \left.  +  2 \frac{(\epsilon_{q_1} k_{n-1}) (\epsilon_{q_2} k_{n-1}) 
(q_1+
 q_2)^2}{(k_{n-1} + 
q_1)^2 
( k_{n-1} +q_1 + q_2)^2 }  \right]
 - 2\frac{(\epsilon_{q_1} k_{n-1}) ( \epsilon_{q_2} k_n) (q_1+ q_2)^2 }{(k_{n-1} 
+ q_1 )^2 (k_n +q_2)^2}
 \right\}  M_{ng} \nonumber \\
\label{now12}
\end{eqnarray}
This expression shows a non-trivial pole structure and the various pole singularities
are all consistent with  the structure of a color-ordered amplitude. In Fig. (\ref{Diagrams}) we sketch  the various pole structures appearing in the previous equation. 
 Notice that the final expression 
in Eq. (\ref{now12}) is symmetric under the simultaneous change of 
$k_{n-1} \leftrightarrow k_{n}$ and $q_1 \leftrightarrow q_2$.

Finally, Eq. (\ref{now}) reproduces Eqs. (2.25) and (2.28) of 
Ref.~\cite{Volovich:2015yoa} in 
the gauge chosen there,
namely $(\epsilon_{q_2} k_n) =(\epsilon_{q_1} k_{n-1}) =0$. 

In the final part of this section we discuss the case of a color-ordered 
amplitude where 
the two soft gluons are not next to each other.  In particular, we consider the 
color-ordered amplitude where a hard gluon, 
say the one with momentum $k_{n-1}$, is between the two soft. 
The gluons are then ordered as 
$k_1 k_2 \dots k_{n-2} \, q_1 \, k_{n-1} \, q_2 \, k_n$.  The complete
amplitude is equal to
\begin{eqnarray}
&& M_{2g;ng}=\frac{(2\alpha')^{\frac{n-2}{2}} }{\alpha'} g_{p+1}^{n-2}  \int
 \frac{ \prod_{i=1}^{n-2} dz_i dz_n }{dV_{abc}} 
\prod_{i=1}^{n} d \theta_i \langle 0 | 
 \prod_{i=1}^{n} {\rm e}^{i \left( \theta_i 
\epsilon \partial_{z_i} +
\sqrt{2\alpha'} k_{i}   \right) X  (z_i) } |0 \rangle  
 (2 \alpha' g_{p+1}^2)  \nonumber \\
&& \times  \int_{0}^{z_{n-2}} dw_1 \int_0^{w_1}dz_{n-1} \int_{0}^{z_{n-1}} dw_2  
\prod_{i=1}^{n-3} (z_i - w_1)^{2\alpha' k_i q_1}\prod_{i=1}^{n-2} 
(z_i - w_2)^{2\alpha'
 k_i q_2} (z_{n-2}-w_1)^{2\alpha' q_1\cdot k_{n-2}} \nonumber \\
&& \times 
(w_1 - z_{n-1} )^{2\alpha' q_1 k_{n-1}} (z_{n-1}-w_2 )^{2\alpha' q_2 k_{n-1}} 
w_1^{2\alpha' q_1\cdot k_n}w_2^{2\alpha' q_2\cdot k_n}
 (w_1- w_2)^{2\alpha' q_1 q_2} \nonumber \\
&& \times  \prod_{i=1}^{n} \left( {\rm e}^{\sqrt{2\alpha'} 
 \frac{\theta_i \epsilon_i q_1}{z_i - w_1}} {\rm e}^{\sqrt{2\alpha'}  
\frac{\theta_i \epsilon_i q_2}{z_i - w_2}} \right)
  \nonumber \\
 &&\times  \left\{   \frac{ (\epsilon_{q_1} \epsilon_{q_2})}{(w_1 - w_2)^2}   
+  \left[  \sum_{i=1}^{n}  \frac{\theta_i ( \epsilon_i  \epsilon_{q_1} 
)}{(z_i - w_1)^2}  - \sum_{i=1}^{n-2} \frac{\sqrt{2\alpha'} (k_i 
\epsilon_{q_1} )}{z_i - w_1}  \right. \right. \nonumber \\
&& \left.  +  \frac{\sqrt{2\alpha'} (k_{n-1} 
\epsilon_{q_1} )}{ w_1 - z_{n-1}}  +   \frac{\sqrt{2\alpha'} (k_n 
\epsilon_{q_1} )}{ w_1 } 
   + \frac{\sqrt{2\alpha'} (\epsilon_{q_1} q_2 )}{w_1 - w_2}    \right] 
\nonumber \\
&& \left.  \times  \left[  \sum_{j=1}^{n} \frac{\theta_j ( 
\epsilon_j \epsilon_{q_2}  )}{(z_j - w_2)^2}  - \sum_{j=1}^{n-1} 
\frac{\sqrt{2\alpha'} (k_j \epsilon_{q_2} )}{z_j - w_2} +  
\frac{\sqrt{2\alpha'} (k_n \epsilon_{q_2} )}{ w_2 } 
  - \frac{\sqrt{2\alpha'} (\epsilon_{q_2} q_1 )}{w_1 - w_2}    \right]  \right\} \ .
\label{soft2gluonsvv}
\end{eqnarray}
Since we are interested in the leading double-soft behavior, we can neglect 
the terms with
$\theta_i$ in the last four lines of the previous expression and keep in 
the curly bracket only the terms
with momenta. It can be seen that the leading double-soft 
behavior is obtained by restricting ourselves to the following expression:
\begin{eqnarray}
M_{2g;ng}= && \frac{(2\alpha')^{\frac{n-2}{2}} }{\alpha'} g_{p+1}^{n-2}  \int
 \frac{ \prod_{i=1}^{n-2} dz_i dz_n }{dV_{abc}} 
\prod_{i=1}^{n} d \theta_i \langle 0 | 
 \prod_{i=1}^{n} {\rm e}^{i \left( \theta_i 
\epsilon \partial_{z_i} +
\sqrt{2\alpha'} k_{i}   \right) X  (z_i) } |0 \rangle  
  \nonumber \\
&& \times  (2 \alpha' )^2 g_{p+1}^2  \int_{0}^{z_{n-2}} 
dw_1 \int_0^{w_1}dz_{n-1} \int_{0}^{z_{n-1}} dw_2  
(z_{n-2}-w_1)^{2\alpha' q_1\cdot k_{n-2}} \nonumber \\
&& \times 
(w_1 - z_{n-1} )^{2\alpha' q_1 k_{n-1}} (z_{n-1}-w_2 )^{2\alpha' q_2 k_{n-1}} 
w_2^{2\alpha' q_2\cdot k_n}
 (w_1- w_2)^{2\alpha' q_1 q_2} \nonumber \\
&& \times \left[  - \frac{ (k_{n-2}  \epsilon_{q_1} )}{  z_{n-2} - w_1} 
  + \frac{ (k_{n-1}  \epsilon_{q_1} )}{ w_1 - z_{n-1}}     
\right]  
\left[    
-  \frac{ (k_{n-1}  \epsilon_{q_2} )}{z_{n-1} - w_2} +  
\frac{ (k_n \epsilon_{q_2} )}{ w_2 }     
\right]  \ .
\label{soft2gluonsvvw}
\end{eqnarray}
In order to study the double-soft behavior of the four terms that appear in 
the last line 
of the previous equation, it is convenient to go to  the  variables 
$x_i$ with
$i=1,2,3$ that run from 0 to 1:
\begin{eqnarray}
w_1 = z_{n-2} x_1 ~~;~~ z_{n-1}= z_{n-2} x_1 x_2~~;~~ w_2 =z_{n-2} x_1 x_2
\label{w1zn-1w2}
\end{eqnarray}
The Jacobian   of the transformation is equal to $z_{n-2}^3 x_1^2 x_2$ and 
the last three lines of Eq.  (\ref{soft2gluonsvvw}) become:
\begin{eqnarray}
&& (2 \alpha' )^2 g_{p+1}^2  \int_0^1 dx_1  \int_0^1 dx_2  \int_0^1 dx_3 
[ z_{n-2} (1-x_1)]^{2\alpha' q_1 k_{n-2}} [ z_{n-2} x_1 (1-x_2)]^{2\alpha' 
q_1 k_{n-1}}
\nonumber \\
&& \times [ z_{n-2} x_1 x_2 (1-x_3)]^{2\alpha' q_2 k_{n-1}} 
[ z_{n-2} x_1 x_2 x_3 ]^{2\alpha' q_2 k_{n}} [ z_{n-2} x_1 (1-x_2 x_3)]^{2
\alpha' q_1 q_2}
\nonumber \\
&& \times  z_{n-2} 
\left[ - x_1  \frac{k_{n-2} \epsilon_{q_1}}{1-x_1} +  \frac{k_{n-1} 
\epsilon_{q_1}}{1-x_2}
\right] \left[ -  \frac{k_{n-1} \epsilon_{q_2}}{1-x_3} +  
\frac{k_{n} \epsilon_{q_2}}{x_3}  \right]
\label{efterchange}
\end{eqnarray}
Let us consider the first term of the square bracket in the last line of the
 previous equation.
The leading double-soft limit is obtained for $x_1 \sim 1$ and $x_3 \sim 1$ 
and $x_3 \sim 0$ for
the two terms in the second square bracket, respectively.  
In this corner of the integration region $z_{n-1}\equiv x_2 z_{n-2}$
and therefore we get
\begin{eqnarray}
&& (2 \alpha' )^2 g_{p+1}^2  z_{n-2} \int_0^1 dx_2 \frac{k_{n-2} \epsilon_{q_1}
}{2\alpha' q_1 k_{n-2}} 
 \left[ \frac{k_{n-1} \epsilon_{q_2}}{2\alpha' q_2 k_{n-1}} -
 \frac{k_n \epsilon_{q_2}
 }{2\alpha' q_2 k_{n}}      \right]  \nonumber \\
 && =
 g_{p+1}^2  \frac{k_{n-2} \epsilon_{q_1}}{ q_1 k_{n-2}} 
 \left[ \frac{k_{n-1} \epsilon_{q_2}}{ q_2 k_{n-1}} - \frac{k_n \epsilon_{q_2}
 }{q_2 k_{n}}      \right]    \int_{0}^{z_{n-2}} d z_{n-1}
\label{foersteto}
\end{eqnarray}
When multiplied with the first line of Eq. (\ref{soft2gluonsvvw}), one gets  the two soft
terms in Eq. (\ref{foersteto}) times the amplitude with $n$ gluons. 

The leading 
double-soft behavior 
of the product of the second terms of the  two square brackets in the last line of 
Eq. (\ref{efterchange}) is obtained for $x_2 \sim 1$ and $x_3 \sim 0$. Therefore, one
gets:
\begin{eqnarray}
(2 \alpha' )^2 g_{p+1}^2  z_{n-2} \int_0^1 dx_1 
\frac{k_{n-1} \epsilon_{q_1}}{2\alpha' k_{n-1} q_1}
\frac{k_{n} \epsilon_{q_2}}{2\alpha' k_{n} q_2}= g_{p+1}^2 \frac{k_{n-1} 
\epsilon_{q_1}}{ k_{n-1} q_1} \frac{k_{n} \epsilon_{q_2}}{ k_{n} q_2}  \int_{0}^{z_{n-2}} d z_{n-1}
\label{andenled}
\end{eqnarray}
 where we have used that, for $x_2 \sim1$, we can write $z_{n-1} = z_{n-2} x_1$.  
 When multiplied with the first line of Eq. (\ref{soft2gluonsvvw}), one gets  again 
 the  soft
term in Eq. (\ref{andenled}) times the amplitude with $n$ gluons. 

The leading double-soft behavior of the last term is a bit more complicated to extract. The 
relevant terms of  Eq. (\ref{efterchange}) that one needs to consider, are:
\begin{eqnarray}
&& - (2 \alpha' )^2 g_{p+1}^2   z_{n-2}  (k_{n-1} \epsilon_{q_1})(k_{n-1} \epsilon_{q_2})
 \int_0^1 dx_1  \int_0^1 dx_2 \,\,  x_2^{2\alpha' (k_{n-1} + k_n)q_2  }
 (1-x_2)^{2\alpha' q_1 k_{n-1}-1}\nonumber \\
&& \times  \int_0^1 dx_3 
 \,\,  (1-x_3)^{2\alpha' q_2 k_{n-1}-1} x_3^{2\alpha' q_2 k_{n}} 
  (1-x_2 x_3)^{2\alpha' q_1 q_2}
\label{sidsteled}
\end{eqnarray}
The dominant contribution comes from the region of the integrals 
around $x_2 \sim 1$ and
$x_3 \sim 1$.  It is convenient to first change variables to 
$y_i = 1 - x_i$ for $i=2,3$ and then 
to $y_2 =tu$ and $y_3 = t (1-u)$ with Jacobian equal to $t$. 
Then, Eq. (\ref{sidsteled}) 
becomes
\begin{eqnarray}
&& - (2 \alpha' )^2 g_{p+1}^2   z_{n-2}  (k_{n-1} \epsilon_{q_1})(k_{n-1} 
\epsilon_{q_2})
 \int_0^1 dx_1  \int_0^{\epsilon} dt \,\, t^{2\alpha' (k_{n-1}(q_1 + q_2) + q_1 q_2  ) -1} 
 \nonumber \\
&& \times \int_0^1 du \,\,  u^{2\alpha' k_{n-1} q_1 -1} (1-u)^{2\alpha' q_2 
k_{n-1}-1} 
  (1-tu)^{2\alpha' q_2 (k_n+k_{n-1})} (1- t(1-u))^{2\alpha' k_n q_2} 
\nonumber \\
&& =   - (2 \alpha' )^2 g_{p+1}^2   z_{n-2}  (k_{n-1} \epsilon_{q_1})(k_{n-1} 
\epsilon_{q_2})
 \int_0^1 dx_1 \frac{1}{2\alpha' (k_{n-1}(q_1 + q_2) + q_1 q_2  ) } 
 \nonumber \\
 && \times\left[ \frac{1}{2\alpha' k_{n-1} q_1} +
 \frac{1}{2\alpha' k_{n-1} q_2}   \right]
\label{sidsteled3}
\end{eqnarray}
where we have neglected irrelevant factors that go to $1$  in the double-soft limit and,
by introducing a cutoff $\epsilon$,  we  have restricted the region of integration for $t$
to a small interval around $t=0$. Eq. (\ref{sidsteled3})  is equal to
\begin{eqnarray}
- g_{p+1}^2      \int_0^{z_{n-2}} d z_{n-1}  
 \frac{ (k_{n-1} \epsilon_{q_1})(k_{n-1} \epsilon_{q_2})}{ 
 (k_{n-1}(q_1 + q_2) + q_1 q_2  ) } \left[ \frac{1}{ k_{n-1} q_1} +
 \frac{1}{ k_{n-1} q_2}   \right]
\label{sidsteb}
\end{eqnarray}
where, for $x_2 \sim 1$, we have used that $ z_{n-1} = z_{n-2} x_1$. 
When we insert the result in Eq. (\ref{sidsteb}) in Eq. (\ref{efterchange}), 
we get again 
the double-soft factor times the amplitude with $n$ gluons.  

In conclusion, the double-soft behavior is given by the sum of the previous
three terms:
\begin{eqnarray}
 M_{2g;ng} &=&  g_{p+1}^2 \left[ \frac{ k_{n-2} \epsilon_{q_1}}{k_{n-2} q_1 } 
\left(  \frac{k_{n-1} \epsilon_{q_2}}{k_{n-1} q_2} -  \frac{k_{n} 
\epsilon_{q_2}}{k_{n} q_2} \right) +  \frac{k_{n-1} \epsilon_{q_1}}{k_{n-1} q_1}
 \frac{k_{n} \epsilon_{q_2}}{k_{n} q_2}  \right. \nonumber \\
   &-&  \left. \frac{ (\epsilon_{q_1} k_{n-1})  (\epsilon_{q_2} k_{n-1}) }{ 
k_{n-1}(q_1 + q_2) + q_1 q_2   } \left( \frac{1}{ k_{n-1} q_1} +
 \frac{1}{ k_{n-1} q_2}   \right)  \right]M_{ng}
\label{sidstesidste} 
\end{eqnarray}  
 It is easy to check that the soft factor is gauge invariant up to terms of order
$q_{1,2}^0$ as in the case of two contiguous soft gluons.

Finally, the double-soft behavior when the two soft gluons are separated
by more than one leg is just the product of two single soft behaviors.

\section{Double-soft behavior with $2$ scalars and $n$ gluons}
\setcounter{equation}{0}
\label{double-soft-with-scalars-gluon}

In this section we compute the double-soft behavior of  two  scalars in an amplitude
with 
$n$ gluons. The amplitude with two scalars and $n$ gluons can be obtained
from that computed in the previous section by noticing that, in this case, we 
have to impose:
\begin{eqnarray}
(\epsilon_i \epsilon_{q_1} )= (\epsilon_i \epsilon_{q_2} ) = (\epsilon_{q_1} k_i ) =
(\epsilon_{q_2} k_i ) = (\epsilon_{q_1} q_2 ) = (\epsilon_{q_2} q_1 )=0  \ .
\label{zerobb}
\end{eqnarray}
 We then get
\begin{eqnarray}
M_{2s; ng} = M_{ng} * H_n
\label{M2sngoo}
\end{eqnarray}
where $M_{ng}$ is given in Eq. (\ref{Annnnn}) and 
\begin{eqnarray}
 H_n =   &&(2 \alpha' g_{p+1}^2)  \int_{0}^{z_{n-1}} dw_1  \int_{0}^{w_1} dw_2  
\prod_{i=1}^{n} \left( |z_i - w_1|^{2\alpha' k_i q_1} |z_i - w_2|^{2\alpha' k_i q_2} 
\right) \nonumber \\
&& \times
 (w_1- w_2)^{2\alpha' q_1 q_2}  \prod_{i=1}^{n} 
\left( {\rm e}^{\sqrt{2\alpha'}  \frac{\theta_i \epsilon_i q_1}{z_i - w_1}}
 {\rm e}^{\sqrt{2\alpha'}  \frac{\theta_i \epsilon_i q_2}{z_i - w_2}} \right)
\frac{ (\epsilon_{q_1} \epsilon_{q_2})}{(w_1 - w_2)^2}   \ .
\label{softfactorSbiss}
\end{eqnarray}
Proceeding as in the previous section
we get the integral in Eq. (\ref{Iagain}) 
computed in Eq. (\ref{symmII}).  Inserting it in the previous equation, we get
\begin{eqnarray} 
H_n =- \frac{g_{p+1}^{2}}{2q_1 q_2} \left[\frac{k_n (q_2 - q_1) + q_1 q_2}{k_{n}
 (q_1 + q_2) + q_1  q_2} + \frac{ k_{n-1} (q_1 - q_2) + q_1 q_2}{k_{n-1}
 (q_1 + q_2) + q_1  q_2}   \right] \ ,
\label{Hnfinal}
\end{eqnarray}
that implies
\begin{eqnarray}
M_{2s;ng} = -\frac{g_{p+1}^{2}}{2q_1 q_2} \left[\frac{k_n (q_2 - q_1) + q_1 q_2}{k_{n}
 (q_1 + q_2) + q_1  q_2} + \frac{ k_{n-1} (q_1 - q_2) + q_1 q_2}{k_{n-1}
 (q_1 + q_2) + q_1  q_2}   \right] M_{ng} \ .
\label{Mng2sofinal}
\end{eqnarray}

\section{Double-soft behavior with $n+2$ scalars}
\setcounter{equation}{0}
\label{double-soft-with-scalars}

In this section we compute the double-soft behavior  of two scalars 
in an amplitude with $n$ additional scalar 
particles instead of gluons. The set-up and the procedure 
is as in the  section for gluons.
The scattering amplitude is given by:
\begin{eqnarray}
 M_{2s;ns} = && \frac{(\sqrt{2\alpha'})^n}{\alpha'} g_{p+1}^{n}
\int \frac{ \prod_{i=1}^{n} dz_i }{dV_{abc}} \prod_{i=1}^{n} d \theta_i 
  \int_0^{z_{n-1}} dw_1 \int_0^{w_1} dw_2 \int d\phi_1 d \phi_2 \nonumber \\
&&\times \langle 0 |  \prod_{i=1}^{n-1} {\rm e}^{i \left( \theta_i \epsilon_i 
\partial_{z_i} +\sqrt{2\alpha'} k_{i}   \right) X (z_i) }    
\label{originalgg} \\
&&  \qquad {\rm e}^{i \left( \phi_1 \epsilon_{q_1}   \partial_{w_1} +
\sqrt{2\alpha'} q_{1}   \right) X (w_1) }   {\rm e}^{i 
\left( \phi_2 \epsilon_{q_2}  \partial_{w_2} +\sqrt{2\alpha'} 
q_{2}   \right) X (w_2) }  {\rm e}^{i \left( \theta_n 
\epsilon_n \partial_{z_n} +\sqrt{2\alpha'} k_{n} 
 \right) X  (z_n) } |0 \rangle \ . \nonumber
\end{eqnarray}
Contracting the vertex operators of the states with momenta $q_1$ and $q_2$
and remembering that, in this case,  all  momenta are orthogonal 
to all polarizations, we get
\begin{eqnarray}
 M_{2s;ns} = && \frac{(\sqrt{2\alpha'})^n}{\alpha'} g_{p+1}^{n}
 \int \frac{ \prod_{i=1}^{n} dz_i }{dV_{abc}} \prod_{i=1}^{n} d \theta_i 
\langle 0 |  \prod_{i=1}^{n} {\rm e}^{i \left( \theta_i \epsilon
\partial_{z_i} +\sqrt{2\alpha'} k_{i} \right) X (z_i) } |0 \rangle   \nonumber \\
&& \times  \int_{0}^{z_{n-1}} dw_1 \int_0^{w_1} dw_2 
 \prod_{i=1}^{n-1} \left( (z_i - w_1)^{2\alpha' k_i q_1} 
(z_i - w_2)^{2\alpha' k_i q_2} \right) \nonumber \\
&&\times(w_1 - z_n )^{2\alpha' q_1 k_n} (w_2 - z_n )^{2\alpha' q_2 k_n}  (w_1- w_2)^{2\alpha' q_1 q_2} 
\nonumber \\
&& \times 
\int d\phi_1 d \phi_2  \,\, {\rm e}^{- \phi_1 \phi_2 
\frac{ (\epsilon_{q_1} \epsilon_{q_2})}{(w_1 - w_2)^2}}
 {\rm e}^{ \phi_1  \sum_{i=1}^{n}  \frac{\theta_i 
(  \epsilon_i  \epsilon_{q_1})}{(z_i - w_1)^2}    } 
 {\rm e}^{ \phi_2  \sum_{i=1}^{n} 
\frac{\theta_i ( \epsilon_i  \epsilon_{q_2} )}{(z_i - w_2)^2}   } \ .
\label{n+2scalar56}
\end{eqnarray}
We can now integrate  over $\phi_1$ and $\phi_2$ 
\begin{eqnarray}
 M_{2s;ns} =&& g_{p+1}^{n-2} \frac{ (\sqrt{2 \alpha'})^{n-2}}{\alpha'} 
\int \frac{ \prod_{i=1}^{n} dz_i }{dV_{abc}} \prod_{i=1}^{n} d \theta_i  
\langle 0 |  \prod_{i=1}^{n} {\rm e}^{i \left( \theta_i \epsilon
\partial_{z_i} +\sqrt{2\alpha'} k_{i}  \right) X  (z_i) } |0 
\rangle  \nonumber \\
&& \times  2 \alpha' g_{p+1}^{2} 
\int_{0}^{z_{n-1}} dw_1 \int_0^{w_1} dw_2  \prod_{i=1}^{n} \left(
(z_i - w_1)^{2\alpha' k_i q_1} (z_i - w_2)^{2\alpha' k_i q_2}  \right) 
\label{m2snsqq}\\
&&  \times (w_1- w_2)^{2\alpha' q_1 q_2}  
 \left[  \frac{(\epsilon_{q_1} \epsilon_{q_2})}{(w_1 - w_2)^2}  +
\sum_{i,j=1}^{n} \frac{\theta_i \theta_j  (\epsilon_{q_1} \epsilon_i) 
(\epsilon_{q_2} \epsilon_j)}{(z_i - w_1)^2 (z_j - w_2)^2} 
\right] \equiv M_{ns} * S_n \ , \nonumber 
\end{eqnarray}
where
\begin{eqnarray}
M_{ns} =g_{p+1}^{n-2} \frac{ (\sqrt{2 \alpha'})^{n-2}}{\alpha'} 
\int \frac{ \prod_{i=1}^{n} dz_i }{dV_{abc}} \prod_{i=1}^{n} d \theta_i 
\langle 0 |  \prod_{i=1}^{n} {\rm e}^{i \left( \theta_i \epsilon_i
\partial_{z_i} +\sqrt{2\alpha'} k_{i} \right) X (z_i) } |0 
\rangle \ ,
\label{Mns}
\end{eqnarray}
 and
\begin{eqnarray}
 S_n =  &&2 \alpha' g_{p+1}^{2}
\int_{0}^{z_{n-1}} dw_1 \int_0^{w_1} dw_2  \prod_{i=1}^{n} \left(
(z_i - w_1)^{2\alpha' k_i q_1} (z_i - w_2)^{2\alpha' k_i q_2}  \right) \nonumber \\
&&  \times (w_1- w_2)^{2\alpha' q_1 q_2}  
 \left[  \frac{(\epsilon_{q_1} \epsilon_{q_2})}{(w_1 - w_2)^2}  +
\sum_{i,j=1}^{n} \frac{\theta_i \theta_j  (\epsilon_{q_1} \epsilon_i) 
(\epsilon_{q_2} \epsilon_j)}{(z_i - w_1)^2 (z_j - w_2)^2} 
\right] \ .
\label{Snnnn}
\end{eqnarray}
Similar to the case of gluons, we keep only the  leading terms 
in the double-soft limit.
It is also convenient to introduce the new variables 
$w_1 = z_{n-1} {\hat{w}}_1$ and 
$w_2 = z_{n-1} {\hat{w}}_2$. Then, the previous expression becomes
 \begin{eqnarray}
 S_n = &&  2 \alpha' g_{p+1}^{2} z_{n-1}^{2\alpha' [ q_1 q_2 + (k_n + k_{n-1})(q_1 + q_2)  ]}
\int_{0}^{1} d{\hat{w}}_1 \int_0^{{\hat{w}}_1} d{\hat{w}}_2  
 \nonumber \\
&& \times
(1- {\hat{w}}_1)^{2\alpha' k_{n-1} q_1} 
(1- {\hat{w}}_2)^{2\alpha' k_{n-1} q_2} 
  {\hat{w}}_{1}^{2\alpha' k_n q_1} {\hat{w}}_{2}^{2\alpha' k_n q_2}  
 ({\hat{w}}_1- {\hat{w}}_2)^{2\alpha' q_1 q_2}   \nonumber \\
&& \times 
 \left[  \frac{(\epsilon_{q_1} \epsilon_{q_2})}{({\hat{w}}_1 - {\hat{w}}_2)^2}  + z_{n-1}^{2}
\sum_{i,j=1}^{n} \frac{\theta_i \theta_j  (\epsilon_{q_1} \epsilon_i) 
(\epsilon_{q_2} \epsilon_j)}{(z_i - z_{n-1} {\hat{w}}_1)^2 (z_j - z_{n-1}{\hat{w}}_2)^2} 
\right] \ .
\label{SnnnnNNN}
\end{eqnarray}
 Also in this case we have two kinds of terms. The first one is without any dependence on the
variables $\theta_i$ and the convolution multiplies the amplitude with $n$ scalars, 
while the second term contains terms with $\theta_i$ that modify the structure of $M_{ns}$. 
Also in this case it can be shown that only the first term contributes to the leading
double-soft limit. Therefore,  here we
concentrate  on the first term taking the term 
\mbox{$z_{n-1}^{2\alpha' [ q_1 q_2 + (k_n + k_{n-1})(q_1 + q_2)  ]}
=1$}. We get the integral in Eq. (\ref{Iagain}) that has been computed in Eq. (\ref{symmII}) and therefore
for the first term in the squared bracket in Eq. (\ref{SnnnnNNN}), one gets:
\begin{eqnarray}
S_n^{(1)} = - \frac{g_{p+1}^{2}}{2q_1 q_2} \left[ \frac{k_n (q_2 - q_1) + q_1 q_2}{k_{n}
 (q_1 + q_2) + q_1  q_2} + \frac{ k_{n-1} (q_1 - q_2) + q_1 q_2}{k_{n-1}
 (q_1 + q_2) + q_1  q_2}   \right] \ .
\label{Sn11}
\end{eqnarray}
This means that the double-soft behavior  in an amplitude with $n+2$ scalars is given by
\begin{eqnarray}
M_{2s;ns} =  -\frac{g_{p+1}^{2}}{2q_1 q_2} \left[ \frac{k_n (q_2 - q_1) + q_1 q_2}{k_{n}
 (q_1 + q_2) + q_1  q_2} + \frac{ k_{n-1} (q_1 - q_2) + q_1 q_2}{k_{n-1}
 (q_1 + q_2) + q_1  q_2}   \right] M_{ns} \ .
\label{M2nsns}
\end{eqnarray}

\section{Conclusions}
\label{conclusions}
\setcounter{equation}{0}

In this paper we have computed the leading double-soft behavior 
of the scattering amplitudes with gluons and scalars living in the world-volume of a 
D$p$-brane of the bosonic string. The corresponding field-theory results can  be obtained by
just 
sending $\alpha'$ to zero.  Our results are valid in any number of space-time dimensions 
and for an arbitrary gauge choice.
We have also for the first time considered and provided the results for  
the double-soft limit of two gluons that are not contiguous,
but separated by a hard gluon in the color-ordered amplitude.

\vspace{-5mm}
\subsection*{Acknowledgments} \vspace{-3mm}
We thank  Massimo Bianchi, Andrea Guerrieri and Congkao Wen   
 for  discussions.

\clearpage

\appendix

\section{Computation of various integrals to  the leading order in the double-soft limit }
\label{computation}

In this Appendix we compute the contribution of various integrals in the double-soft limit.

Let us start considering the first integral appearingÊ inÊ Eq.Ê (\ref{Snalmostcomple1}):

\begin{eqnarray}
I (k_2 , q_1 ; k_3 , q_2)  \equiv&& \int_{0}^{1} d w_1 \int_{0}^{w_1} dw_2  
w_1^{2 \alpha' q_1 k_3} w_2^{2 \alpha' q_2 k_3} (1-w_1)^{2\alpha'k_2 q_1} 
(1-w_2)^{2\alpha'k_2 q_2}  \nonumber \\
&& \times  (w_1 - w_2)^{2 \alpha' q_1 q_2 -2}
\label{Iagain}
\end{eqnarray}
and let us show that it is symmetric under the simultaneous exchange of $k_2 \leftrightarrow k_3$ and
$q_1 \leftrightarrow q_2$. It is convenient to introduce the variables:
\begin{eqnarray}
x_4\equiv w_2~~;~~x_3\equiv 1-w_1
\label{var}
\end{eqnarray}
The integral becomes:
\begin{eqnarray}
I (k_2 , q_1 ; k_3 , q_2) = &&\int_{0}^{1} dx_3 (1-x_3)^{2 \alpha' q_1 k_3} x_3^{2\alpha' q_1 k_2}
 \int_{0}^{1-x_3} dx_4 x_4^{2\alpha' q_2 k_3} (1-x_4)^{2\alpha' k_2 q_2}  \nonumber \\
&&\times  (1-x_3 -x_4)^{2\alpha' q_1 q_2 -2}
\label{2soft3tac1}
\end{eqnarray} 
It is possible to give an equivalent expression of the integral (\ref{Iagain}):
\begin{eqnarray}
I (k_2 , q_1 ; k_3 , q_2)=&&\int_{0}^{1} dw_2  w_2^{2\alpha' q_2 k_3}
 (1-w_2)^{2\alpha' k_2 q_2} \int_{w_2}^1dw_1
w_1^{2 \alpha' q_1 k_3} (1-w_1)^{2\alpha' q_1 k_2}   \nonumber \\
&& \times (w_1 - w_2)^{2\alpha' q_1 q_2 -2}
\label{2soft3tac2}
\end{eqnarray}
Introducing again the variables in Eq. (\ref{var}), we get
\begin{eqnarray}
I (k_2 , q_1 ; k_3 , q_2)
=&&\int_{0}^{1} dx_4  x_4^{2\alpha' q_2 k_3} (1-x_4)^{2\alpha' k_2 q_2} \int_{0}^{1-x_4}dx_3
(1-x_3)^{2 \alpha' q_1 k_3} x_3^{2\alpha' q_1 k_2} \nonumber \\
&& \times (1-x_3 - x_4)^{2\alpha' q_1 q_2 -2}\nonumber\\
\label{2soft3tac2b}
\end{eqnarray}
Both representations can be collected in the formula:
\begin{eqnarray}
I (k_2 , q_1 ; k_3 , q_2)=&&\int_0^1\prod_{i=1}^3dx_i 
\delta(1-x_2-x_3 -x_4 )x_4^{2\alpha' q_2 k_3} (1-x_4)^{2\alpha' 
k_2 q_2} (1-x_3)^{2 \alpha' q_1 k_3}  \nonumber \\
&& \times x_3^{2\alpha' q_1 k_2}  x_2^{2\alpha' q_1 q_2 -2}
\end{eqnarray}
Let  us now keep the variable $x_2$ and  solve the delta function by  
eliminating  either $x_3$ or $x_4$.  By eliminating $x_3=1-x_2-x_4$ one gets:
\begin{eqnarray}
I (k_2 , q_1 ; k_3 , q_2)= &&
\int_0^1dx_2\int_0^{1-x_2} d x_4 x_4^{2\alpha' q_2 k_3} (1-x_4)^{2\alpha' k_2 q_2}
\nonumber \\
&& \times
(x_2+x_4)^{2 \alpha' q_1 k_3} (1-x_2-x_4)^{2\alpha' q_1 k_2}  x_2^{2\alpha' q_1 q_2 -2}
\end{eqnarray}
Introducing the variables $x_2=\rho \omega~~;~~x_4=\rho(1-\omega)$ one gets:
\begin{eqnarray}
I (k_2 , q_1 ; k_3 , q_2)=&&\int_0^1d\rho  \rho \int_0^{1}d\omega  \rho^{2\alpha' q_2 k_3} (1-\omega)^{2\alpha' q_2 k_3} (1-\rho(1-\omega))^{2\alpha' k_2 q_2} \nonumber \\
&& \times \rho^{2 \alpha' q_1 k_3} (1-\rho)^{2\alpha' q_1 k_2}  \rho^{2\alpha' q_1 q_2 -2}\omega^{2\alpha' q_1 q_2 -2}
\label{1.71}
\end{eqnarray}
By changing variable $t=1-\omega$, one gets:
\begin{eqnarray}
I (k_2 , q_1 ; k_3 , q_2)
=&&\int_0^1d\rho \rho^{2\alpha'(q_2+q_1)\cdot k_3+2\alpha' q_1 q_2 -1}(1-\rho)^{2\alpha' q_1 k_2}\int_0^{1}dt   t^{2\alpha' q_2 k_3}  \nonumber \\
&& \times (1-\rho~t)^{2\alpha' k_2 q_2}   (1-t)^{2\alpha' q_1 q_2 -2}\nonumber\\
\label{fin}
\end{eqnarray}
%This  expression coincides with Eq. (\ref{II63}).
Let us now eliminate instead $x_4=1-x_2-x_3$.  One gets:
\begin{eqnarray}
I (k_2 , q_1 ; k_3 , q_2)=
&&\int_0^1dx_2\int_0^{1-x_2}dx_3(1-x_2-x_3)^{2\alpha' q_2 k_3} (x_2+x_3)^{2\alpha' k_2 q_2}
(1-x_3)^{2 \alpha' q_1 k_3} \nonumber \\
&& \times x_3^{2\alpha' q_1 k_2}  x_2^{2\alpha' q_1 q_2 -2}
\label{x4instead}
\end{eqnarray}
By defining again $x_2=\rho \omega~~;~~x_3=\rho(1-\omega)$ one gets:
\begin{eqnarray}
I (k_2 , q_1 ; k_3 , q_2)=&&\int_0^1d\rho \rho \int_0^{1}d\omega(1-\rho)^{2\alpha' q_2 k_3} \rho^{2\alpha' k_2 q_2}
(1-\rho(1-\omega))^{2 \alpha' q_1 k_3} \rho^{2\alpha' q_1 k_2}
\nonumber \\
&& \times (1-\omega)^{2\alpha' q_1 k_2}  \rho^{2\alpha' q_1 q_2 -2}\omega^{2\alpha' q_1 q_2 -2}
\end{eqnarray}
This  equation is obtained from Eq. (\ref{1.71}) by changing $q_1\leftrightarrow q_2$ and 
$k_2\leftrightarrow k_3$ and we get
\begin{eqnarray}
I (k_2 , q_1 ; k_3 , q_2)= I (k_3, q_2 ; k_2 , q_1)
\label{symmetryvvv}
\end{eqnarray}  
Let us now compute it in the double-soft limit.
Let us first change variables $w_1 =z$ and  $w_2 = x w_1$ in Eq. (\ref{Iagain}) getting
\begin{eqnarray}
I = && \int_0^1 dz  z^{2\alpha' ( k_3 (q_1 + q_2) + q_1 q_2)-1} 
(1-z)^{2\alpha' k_2 q_1} 
\int_0^1 dx  x^{2 \alpha' q_2 k_3} (1-x)^{2\alpha' q_1 q_2 -2} 
(1-xz)^{2\alpha' k_2 q_2} \nonumber \\
\label{II63bb}
\end{eqnarray}
Then, we use  the following equation
\begin{eqnarray}
\int_0^1 dt  \,\,t^{b-1} (1-t)^{c-b-1} (1- t y)^{-a} = \frac{\Gamma (b) \Gamma (c-b)}{\Gamma (c)} \,\,
{}_2 F_1 (a,b;c;y) 
\label{2F1y}
\end{eqnarray}
to rewrite Eq. (\ref{II63bb}) as follows
\begin{eqnarray}
I = \frac{\Gamma (b) \Gamma (c-b)}{\Gamma (c)}  \int_0^1 dz  z^{2\alpha' ( k_3 (q_1 + q_2) + q_1 q_2)-1} 
(1-z)^{2\alpha' k_2 q_1}  {}_2 F_1 (a,b;c;z)
\label{IIag}
\end{eqnarray}
with
\begin{eqnarray}
b =1+ 2\alpha' q_2 k_3~~&&;~~a= - 2\alpha' k_2 q_2
\nonumber \\
c-b = 2 \alpha' q_1 q_2 -1 
~~&&;~~
c = 2\alpha' (q_1+k_3)q_2 
\label{abcuu}
\end{eqnarray}
Then we can use 
\begin{eqnarray}
{}_2 F_1 (a,b;cz)  = (1-z)^{-a} {}_2 F_1 \left (c-b,a;c; \frac{z}{1-z} \right)
\label{F21for}
\end{eqnarray}
to get
\begin{eqnarray}
 I = &&\frac{\Gamma (b) \Gamma (c-b)}{\Gamma (a) \Gamma (c-a)} \int_0^1 dz  z^{2\alpha' ( k_3 (q_1 + q_2) + q_1 q_2)-1} 
(1-z)^{2\alpha' k_2 (q_1 +q_2) } \nonumber \\
&& \times \int_0^1 dx x^{a-1} (1-x)^{c-a-1 } ( 1 - \frac{zx}{z-1})^{b-c }  \nonumber \\
=&&   \frac{\Gamma (b) \Gamma (c-b)}{\Gamma (a) \Gamma (c-a)} \int_0^1 dz  z^{2\alpha' ( k_3 (q_1 + q_2) + q_1 q_2)-1} 
(1-z)^{2\alpha' k_2 (q_1 +q_2) } \nonumber \\
&& \times \int_0^1 dt (1-t)^{a-1} t^{c-a-1 } \left( \frac{1-zt}{1-z} \right)^{b-c } 
\label{III5s}
\end{eqnarray}
It is equal to
\begin{eqnarray}
I =&&  \frac{\Gamma (b) \Gamma (c-b)}{\Gamma (a) \Gamma (c-a)} \int_0^1 dz  z^{2\alpha' 
( k_3 (q_1 + q_2) + q_1 q_2)-1} 
(1-z)^{2\alpha' (k_2 (q_1 +q_2) + q_1 q_2) -1 } \nonumber \\
&& \times  \int_0^1 dt (1-t)^{- 2 \alpha' k_2 q_2-1} t^{2\alpha' q_2 (k_2 + k_3 +q_1)-1 } 
\left( 1-zt \right)^{1- 2\alpha' q_1 q_2 } 
\label{gggh}
\end{eqnarray}
Introducing, for the sake of simplicity, the momentum $k_1 = - k_2 - k_3 - q_1 - q_2$, 
yields
\begin{eqnarray}
I =&& \frac{\Gamma (b) \Gamma (2\alpha' q_1 q_2 -1)}{\Gamma ( - 2 \alpha' k_2 q_2) 
\Gamma (- 2\alpha' q_2 k_1 )}  \int_0^1 dz  z^{2\alpha' 
( k_3 (q_1 + q_2) + q_1 q_2)-1} 
(1-z)^{2\alpha' (k_2 (q_1 +q_2) + q_1 q_2) -1 } \nonumber \\
&& \times  \int_0^1 
dt (1-t)^{- 2 \alpha' k_2 q_2-1}  t^{-2\alpha' q_2 k_1 -1} \left( 1-zt \right)^{1- 2\alpha' q_1 q_2 }
\label{II764}
\end{eqnarray}
Keeping only the term with 1 in the exponent of the last term, we can easily compute the two integrals.
One gets:
\begin{eqnarray}
I = &&\frac{\Gamma (b) \Gamma (2\alpha' q_1 q_2 -1)}{\Gamma ( - 2 \alpha' k_2 q_2) 
\Gamma (- 2\alpha' q_2 k_1 )}  \left[  \frac{\Gamma ( 2\alpha' 
( k_3 (q_1 + q_2) + q_1 q_2)  ) \Gamma (2\alpha' 
( k_2 (q_1 + q_2) + q_1 q_2) )  }{\Gamma (2\alpha' 
( (k_3 +k_2)  (q_1 + q_2) + 2q_1 q_2) ) }  \right. \nonumber \\
&& \times \frac{ \Gamma (-2\alpha' k_2 q_2 ) \Gamma (- 2\alpha' q_2 k_1 )}{
\Gamma ( - 2 \alpha ' q_2 (k_1 + k_2) )} 
-  \frac{\Gamma ( 2\alpha' 
( k_3 (q_1 + q_2) + q_1 q_2) +1 ) \Gamma (2\alpha' 
( k_2 (q_1 + q_2) + q_1 q_2) )  }{\Gamma (2\alpha' 
( (k_3 +k_2)  (q_1 + q_2) + 2q_1 q_2)+1 )  }  \nonumber \\ 
&& \left. \times   \frac{ \Gamma (-2\alpha' k_2 q_2 ) \Gamma (1- 2\alpha' q_2 k_1 )}{
\Gamma (1 - 2 \alpha ' q_2 (k_1 + k_2) )}  \right]
\label{Iquasiquer}
\end{eqnarray}
It can be rewritten as follows
\begin{eqnarray}
I = &&\frac{\Gamma (b) \Gamma (2\alpha' q_1 q_2 +1)}{ (2\alpha ' q_1 q_2) (2\alpha ' q_1 q_2-1)
\Gamma (- 2\alpha' q_2 k_1 )}  \frac{  
\Gamma (1- 2\alpha' q_2 k_1 )}{
\Gamma (1 - 2 \alpha ' q_2 (k_1 + k_2) )}   \nonumber \\
 && \times  \frac{\Gamma ( 1+ 2\alpha' 
( k_3 (q_1 + q_2) + q_1 q_2)  ) \Gamma ( 1+2\alpha' 
( k_2 (q_1 + q_2) + q_1 q_2) )  }{\Gamma ( 1+2\alpha' 
( (k_3 +k_2)  (q_1 + q_2) + 2q_1 q_2) ) }   \nonumber \\
&& \times \left[  \frac{q_2 (k_1 +k_2)}{ 2\alpha' q_2 k_1} 
\left( \frac{1}{  (k_3 (q_1 + q_2) + q_1 q_2)} +  
\frac{1}{ (k_2 (q_1 + q_2) + q_1 q_2)} \right)  \right. \nonumber \\
&&-  \left. \frac{1}{ 2\alpha' (k_2 (q_1 + q_2) + q_1 q_2)} 
 \right]
\label{Iquasiqua}
\end{eqnarray}
that is equal to
\begin{eqnarray}
I = &&\frac{\Gamma (b)  (-1)\Gamma (2\alpha' q_1 q_2 +1)}{ (2\alpha ' q_1 q_2) (2\alpha ' q_1 q_2-1)
\Gamma (1- 2\alpha' q_2 k_1 )}  \frac{  
\Gamma (1- 2\alpha' q_2 k_1 )}{
\Gamma (1 - 2 \alpha ' q_2 (k_1 + k_2) )}   \nonumber \\
 && \times  \frac{\Gamma ( 1+ 2\alpha' 
( k_3 (q_1 + q_2) + q_1 q_2)  ) \Gamma ( 1+2\alpha' 
( k_2 (q_1 + q_2) + q_1 q_2) )  }{\Gamma ( 1+2\alpha' 
( (k_3 +k_2)  (q_1 + q_2) + 2q_1 q_2) ) }   \nonumber \\
&& \times \left[ \frac{q_2 (k_1 +k_2)}{ (k_3 (q_1 + q_2) + q_1 q_2)} + 
\frac{k_2 q_2  }{  k_2 (q_1 + q_2) + q_1 q_2}     \right]
\label{Ifinaliss}
\end{eqnarray}
In the double-soft  limit one gets:
\begin{eqnarray}
I^{ft}  = \frac{1}{2\alpha' q_1 q_2} \left[  -\frac{q_2 (k_3 + q_1)}{ k_3 (q_1 + q_2) + q_1 q_2} + 
\frac{k_2 q_2  }{  k_2 (q_1 + q_2) + q_1 q_2}    \right]
\label{Ifieldthe}
\end{eqnarray}
Because of the symmetry in Eq. (\ref{symmetryvvv}), we must symmetrize the 
previous expression getting:
\begin{eqnarray}
I^{ft} = -\frac{1}{4\alpha' q_1 q_2} \left[ \frac{k_3 (q_2 - q_1) + q_1 q_2}{k_{3}
 (q_1 + q_2) + q_1  q_2} + \frac{ k_2 (q_1 - q_2) + q_1 q_2}{k_{2}
 (q_1 + q_2) + q_1  q_2}  \right]
\label{symmII}
\end{eqnarray}

In the second part of this section we compute the remaining nine integrals in Eq. (\ref{Snalmostcomple})  in the  double-soft limit ($q_{1}$ 
and $q_{2}$ going simultaneously to zero).

The term with the denominator $\frac{1}{w_1 (w_1-w_2)}$ can be written as 
follows after having taken $w_2 = w_1 x$:
\begin{eqnarray}
 I_{w_1 ; w_1 -w_2} \equiv&& \int_0^1 dw_1 \int_0^{1} dx \,\, 
w_1^{ 2 \alpha' ( k_3(q_1 + q_2) + q_1 q_2) -1} 
(1 - x)^{2 \alpha' q_1 q_2 -1} (1 - w_1)^{2\alpha' k_2 q_1}
   \nonumber \\
&& \times  (1 - w_1 x )^{2\alpha' k_2 q_2}  
w_1^{2\alpha' q_1 k_3} x^{2\alpha' q_2 k_3}  
\nonumber \\
 \sim&&
\frac{1}{ (2\alpha' q_1 q_2 ) (2 \alpha' ( k_3(q_1 + q_2) + q_1 q_2) }
\label{w1(w1-w2)}
\end{eqnarray}
The term with the denominator $\frac{1}{w_2 (w_1-w_2)}$ can be written as 
follows after having taken $w_2 = w_1 x$:
\begin{eqnarray}
I_{w_2 ; w_1 -w_2} \equiv &&   \int_0^1 dw_1 \int_0^{1} dx  \,\,  
 w_1^{2\alpha' ( (q_1+ q_2) k_3 + q_1 q_2) -1}  (1 - x)^{2 \alpha' q_1 q_2 -1} 
   \nonumber \\
&& \times  (1 - w_1 x)^{2\alpha' k_2 q_2}  (1 - w_1)^{2\alpha' k_2 q_1}
  x^{2\alpha' q_2 k_3 -1}   
  \nonumber \\
  \sim&& \frac{1}{2\alpha' ( (q_1+ q_2) k_3 + q_1 q_2) } 
 \frac{\Gamma (2\alpha' k_3 q_2)  
\Gamma ( 2\alpha' q_1 q_2 ) }{\Gamma ( 2\alpha' ( q_1 q_2 + k_3 q_2) }  
\nonumber \\
 \sim &&
\frac{1}{2\alpha' ( (q_1+ q_2) k_3 + q_1 q_2) }  
\frac{2\alpha' (q_1 q_2 + k_3 q_2  ) }{(2\alpha' q_1 q_2 ) (2\alpha'  k_3 q_2 ) }
 \nonumber \\
=&& \frac{1}{ (2\alpha' q_1 q_2 ) (2 \alpha' ( k_3(q_1 + q_2) + q_1 q_2) }
\label{w2(w1-w2)}
\end{eqnarray}
The term with the denominator $\frac{1}{(1 - w_2) (w_1-w_2)}$ can be written as 
follows:
\begin{eqnarray}
  I_{1-w_2 ; w_1 -w_2}  \equiv&& \int_0^1 dw_2 \int_{w_2}^1  dw_2 
\,\, (w_1 - w_2)^{2 \alpha' q_1 q_2-1}
(1 - w_1)^{2\alpha' k_2 q_1}
  \nonumber \\
&& \times   (1 - w_2)^{2\alpha' k_2 q_2 -1} w_1^{2\alpha' q_1 k_3} 
w_2^{2\alpha' q_2 k_3}  
\label{435}
\end{eqnarray}
Introducing the new variables  $w_1 = 1-x_1$ and $w_2 =1-x_2$, one gets
\begin{eqnarray}
&&  I_{1-w_2 ; w_1 -w_2}  \equiv 
\int_0^1 dx_2 \int_{0}^{x_2}   dx_1 \,\, (x_2 - x_1)^{2 \alpha' q_1 q_2-1}
x_1^{2\alpha' k_2 q_1}
  x_2^{2\alpha' k_2 q_2 -1}  \nonumber \\
&& \times (1 -x_1 )^{2\alpha' q_1 k_3} (1- x_2 )^{2\alpha' q_2 k_3}  
\label{436}
\end{eqnarray}
By introducing the variable $x_1 = x_2 x$ we get
\begin{eqnarray}
 I_{1-w_2 ; w_1 -w_2}  \equiv  &&
\int_0^1 dx_2 \int_{0}^{1}   dx  \,\,  x_2^{2\alpha' (q_1 q_2 + k_2 (q_1 +q_2))-1  }
(1 - x)^{2 \alpha' q_1 q_2-1}
x^{2\alpha' k_2 q_1}
   \nonumber \\
&& \times  x_2^{2\alpha' k_2 q_2 -1} 
(1 -x_2 x )^{2\alpha' q_1 k_3} (1- x_2 )^{2\alpha' q_2 k_3}  \nonumber \\ 
\sim && \frac{1}{(2\alpha' q_1 q_2) 2\alpha' ( (q_1 q_2 + k_2 (q_1 +q_2))}
\label{4368}
\end{eqnarray}
The term with the denominator $\frac{1}{(1 - w_1) (w_1-w_2)}$ can be written as:
\begin{eqnarray}
 I_{1-w_1 ; w_1 -w_2}  \equiv  && 
\int_0^1 dw_2 \int_{w_2}^{1} dw_1         (w_1 - w_2)^{2 \alpha' q_1 q_2 -1}
(1 - w_1)^{2\alpha' k_2 q_1-1}
  \nonumber \\
&& \times   (1 - w_2)^{2\alpha' k_2 q_2}  
w_1^{2\alpha' q_1 k_3} w_2^{2\alpha' q_2 k_3}   
\label{1-w1}
\end{eqnarray}
Introducing the variables $w_1 = 1-x_1$ and $w_2 =1-x_2$, one gets:
\begin{eqnarray}
   I_{1-w_1 ; w_1 -w_2}  \equiv  &&
\int_0^1 dx_2 \int_{0}^{x_2} dx_1         (x_2 - x_1)^{2 \alpha' q_1 q_2 -1}
x_1^{2\alpha' k_2 q_1-1}
  x_2^{2\alpha' k_2 q_2}  \nonumber \\
&& \times (1-x_1)^{2\alpha' q_1 k_3} (1-x_2)^{2\alpha' q_2 k_3}   
\label{chvari}
\end{eqnarray}
Changing variable $ x_1 = x_2 x$, we get
\begin{eqnarray}
  I_{1-w_1 ; w_1 -w_2}  \equiv  &&
\int_0^1 dx_2 \int_{0}^{1} dx\,\,  x_2^{2\alpha' ( k_2 (q_1 + q_2) + q_1 q_2)-1}         
(1 - x)^{2 \alpha' q_1 q_2 -1}
  \nonumber \\
&& \times  x^{2\alpha' k_2 q_1-1}  (1-x_2 x)^{2\alpha' q_1 k_3} 
(1-x_2)^{2\alpha' q_2 k_3}   
\nonumber \\
\sim&& \frac{1}{2\alpha' (  k_2 (q_1 + q_2) + q_1 q_2)  )} 
  \frac{\Gamma ( 2\alpha' q_1 q_2) \Gamma (2\alpha' k_2 
q_1)}{\Gamma ( 2\alpha' (q_1 (q_2 + k_2)} 
\nonumber \\
\sim &&
\frac{1}{2\alpha' (  k_2 (q_1 + q_2) + q_1 q_2)  )} \frac{ 2\alpha'( k_2(q_1 + q_2)) }{(2\alpha' q_1 q_2) (2\alpha' k_2 q_1) }
\nonumber \\
= && \frac{1}{2\alpha' (  k_2 (q_1 + q_2) + q_1 q_2)  )} \frac{1}{2\alpha' q_1 q_2}
\label{1-w1final}
\end{eqnarray}
After the
change of variable $w_2 = w_1 x$, the term with the denominator $\frac{1}{w_1 w_2}$
is given by:
\begin{eqnarray}
 I_{w_1; w_2} \equiv&& \int_0^1 dw_1  w_1^{2\alpha' ( k_3 (q_1 + q_2) + q_1 q_2) -1} 
(1-w_1)^{2\alpha' k_2 q_1} \int_0^1 dx x^{2\alpha' q_2 k_3 -1}   
\nonumber \\ 
&& \times (1-x)^{2\alpha' q_1 q_2} (1-xw_1)^{2\alpha' k_2 q_2} 
\nonumber \\
\sim&& \frac{1}{2\alpha' (k_3 (q_1 + q_2) + q_1 q_2)
(2\alpha' k_3 q_2)} 
\label{w1w2nn}
\end{eqnarray}
The term with the denominator $\frac{1}{(1-w_1)(1-w_2)}$ after the change of variables 
$w_1 =1-x_1$ and  $w_2 = 1 - x_2$ is equal to:
\begin{eqnarray}
 I_{1-w_1;1-w_2} \equiv &&
\int_0^1 dx_1   x_1^{2\alpha'  q_1 k_2 -1} (1- x_1)^{2\alpha' q_1 k_3}  
\int_{x_1}^{1} dx_2         
(1 - x_2 )^{2 \alpha'  q_2 k_3 }
x_2^{2\alpha' k_2 q_2-1}   
(x_2-x_1 )^{2\alpha' q_2 q_1} 
\nonumber \\
=&& \int_0^1 dx_2 x_2^{2\alpha' k_2 q_2-1}  
(1-x_2)^{2 \alpha'  q_2 k_3 } 
\int_0^{x_2} dx_1  x_1^{2\alpha'  q_1 k_2 -1} (1- x_1)^{2\alpha' q_1 k_3} 
(x_2-x_1 )^{2\alpha' q_2 q_1}
\nonumber\\ 
\label{1-w1777}
\end{eqnarray} 
After changing variable $ x_1 = x_2 x$ we get
\begin{eqnarray}
I_{1-w_1;1-w_2} \equiv  &&
 \int_0^1 dx_2 x_2^{2\alpha' (k_2 (q_1 + q_2) + q_1 q_2)-1}  
(1-x_2)^{2 \alpha'  q_2 k_3 }  \int_0^1 dx  x^{2\alpha' q_1 k_2-1  } 
\nonumber \\
&& \times  (1-x)^{2\alpha' q_1 q_2}(1 - x_2 x)^{2\alpha' q_1 k_3} 
\nonumber \\
\sim&& \frac{1}{2\alpha' ( k_2 (q_1 + q_2) + q_1 q_2) 
(2\alpha' q_1 k_2)} 
\label{gvfd}
\end{eqnarray} 
It is obtained from the one in Eq. (\ref{w1w2nn}) with the substitution 
$k_3 \leftrightarrow k_2$ and  $q_1 \leftrightarrow q_2$. 

 The term with  the factor $(1-w_1) w_2$ in the denominator is given by
\begin{eqnarray}
I_{1-w_1 ;w_2} \equiv && \int_0^1 dw_1 \int_0^{w_1} dw_2 \   
w_1^{2\alpha' k_3 q_1} (1-w_1)^{2\alpha' k_2 q_1 -1 }
\nonumber \\
&& \times
w_2^{2\alpha' k_3 q_2 -1} (1-w_2)^{2\alpha' k_2 q_2} 
 (w_1 - w_2)^{2\alpha' q_1 q_2} 
\nonumber \\
 = &&  \int_0^1 dw_1 w_1^{2\alpha'  
(k_3 (q_1 +q_2) + q_1 q_2 ) } (1-w_1)^{2\alpha' k_2 q_1 -1 } 
\nonumber \\
&& \times   \int_0^1 dt \,\, t^{2\alpha' k_3 q_2 -1} (1-t)^{2\alpha' 
q_1 q_2} (1- t w_1)^{2\alpha' k_2 q_2  } \nonumber \\
 \sim&& \frac{1}{(2\alpha'k_2 q_1) (2\alpha' k_3 q_2)}
\label{byr}
\end{eqnarray}
where we have kept only the first term  (1) of the expansion 
of the last term in the previous equation. The last integral to compute is the one 
with the terms $w_1 (1-w_2)$ in the 
denominator:
\begin{eqnarray}
 I_{w_1;  1-w_2 } \equiv&&  \int_0^1dw_1\int_0^{w_1}
dw_2(1-w_1)^{2\alpha' q_1 k_2}w_1^{2\alpha'q_1
 k_3-1}(w_1-w_2)^{2\alpha'q_1  q_2} 
 (1-w_2)^{2\alpha'q_2  k_2-1} 
w_2^{2\alpha'q_2  k_3} 
\nonumber \\
=&&\int_0^1dw_1\int_0^{w_1}dw_2
(1-w_1)^{2\alpha' q_1  k_2}w_1^{2\alpha'q_1 k_3-1}
\nonumber \\
&& \times  
(w_1-w_2)^{2\alpha'q_1 q_2} \left(-\frac{1}{2\alpha' q_2 k_2}
\frac{\partial}{\partial w_2}\right)
(1-w_2)^{2\alpha'q_2  k_2} w_2^{2\alpha'q_2 k_3} 
  \nonumber \\
  =&&\int_0^1dw_1\int_0^{w_1}dw_2   (1-w_1)^{2\alpha' q_1 
k_2}w_1^{2\alpha'q_1 k_3-1}(w_1-w_2)^{2\alpha'q_1 
q_2}
\nonumber\\
&&\times
(1-w_2)^{2\alpha'q_2 k_2} w_2^{2\alpha'q_2 k_3} \biggl[\frac{q_2\
\cdot k_3}{q_2  k_2}w_2^{-1}-\frac{q_1 q_2}{q_2
k_2}(w_1-w_2)^{-1}\biggr]= 0
\label{w1(w1-w2)x}
\end{eqnarray}
as one can see from Eqs. (\ref{w1w2nn}) and (\ref{w1(w1-w2)}). Finally, the term 
involving $I_{w_1 - w_2; w_1 -w_2}$ is directly seen to be subleading in the double-soft limit, due to the explicit factor of $q_1 q_2$ that multiplies this integral, and is thus not needed.

\end{document}